\documentclass[acmsmall]{acmart}
\AtBeginDocument{%
  }

\setcopyright{cc}
\setcctype{by}
\acmDOI{10.1145/3832093}
\acmYear{2026}
\acmJournal{PACMSE}
\acmVolume{3}
\acmNumber{ISSTA}
\acmArticle{ISSTA002}
\acmMonth{10}
\acmSubmissionID{issta26main-p17-p}
\received{2026-01-30}
\received[accepted]{2026-06-25}
\makeatletter
\edef\@acmArticle{\noexpand\mbox{\@acmArticle}}
\makeatother

\usepackage{tabularx} 
\AtBeginDocument{%
  }
\usepackage{multirow}
\usepackage{booktabs}
\usepackage{array}
\usepackage{threeparttable}
\usepackage{subcaption}
\usepackage{xcolor}
\usepackage{soul}

\usepackage{algorithmic}
\usepackage{graphicx}
\usepackage{textcomp}
\usepackage{tabularx}
\usepackage{array}
\usepackage{url}
\usepackage{enumitem}
\usepackage{listings}
\usepackage{pifont} 
\usepackage[most]{tcolorbox}
\lstdefinestyle{promptstyle}{
    backgroundcolor=\color{white},    
    frame=single,                         
    breaklines=true,                    
    breakatwhitespace=true,             
    breakindent=10pt,                   
    basicstyle=\scriptsize\ttfamily,  
    showstringspaces=false,             
    keepspaces=true,                    
    numberstyle=\tiny\color{black!50},  
    stepnumber=1,
    numbersep=5pt,
    tabsize=2,
}
\lstdefinestyle{codeexample}{
  language=Java,
  basicstyle=\ttfamily\scriptsize,
  columns=fullflexible,
  keepspaces=true,
  showstringspaces=false,
  frame=none,
  aboveskip=0pt,
  belowskip=0pt,
  breaklines=true,
  breakatwhitespace=false,
  breakindent=0pt,
  xleftmargin=0pt,
  xrightmargin=0pt
}

\usepackage{lstlinebgrd}          

\makeatletter
\let\old@lst@AddFrom\lst@AddFrom
\expandafter\def\expandafter\lst@AddFrom\expandafter#\expandafter1\expandafter#%
  \expandafter2\expandafter#\expandafter3{\old@lst@AddFrom{#1}{#2}{#3}}
\makeatother

\lstdefinestyle{codeSmall}{
  style=codeexample,
  basicstyle=\ttfamily\tiny,
  aboveskip=2pt, belowskip=2pt,
}
\lstdefinestyle{codeTiny}{
  style=codeexample,
  basicstyle=\ttfamily\tiny,
  aboveskip=0pt,
  belowskip=0pt,
}

\tcbset{
  compactfulltitle/.style={
    enhanced,
    sharp corners,
    boxrule=0.4pt,
    colback=white,
    fonttitle=\bfseries\scriptsize,
    coltitle=white,
    left=0pt, right=0pt,
    top=0pt, bottom=0pt,
    toptitle=-0.5pt,
    bottomtitle=-1pt,
    lefttitle=0.5pt,
    righttitle=0.5pt,
  },
  fixedbox/.style={
    compactfulltitle,
    colframe=teal!60!black,
    colbacktitle=teal!70!black,
  },
  buggybox/.style={
    compactfulltitle,
    colframe=red!60!black,
    colbacktitle=red!70!black,
  },
  docstringbox/.style={
    compactfulltitle,
    colframe=orange!70!black,
    colbacktitle=orange!75!black,
  },
}

\newcommand{\hlRange}[2]{%
  \ifnum\value{lstnumber}<#1 \else
  \ifnum\value{lstnumber}>#2 \else
    \color{yellow!30}%
  \fi\fi}

\newcommand{\hltok}[1]{%
  \begingroup
  \setlength{\fboxsep}{0.5pt}%
  \colorbox{yellow!30}{#1}%
  \endgroup
}

\newlength{\methodboxheight}
\newlength{\testboxheight}
\soulregister{\cite}{7}
\soulregister{\citep}{7}
\soulregister{\citeauthor}{7}
\soulregister{\footnote}{7}
\soulregister{\ref}{7}
\soulregister{\texttt}{7}
\soulregister{\textcircled}{7}

\begin{document}

\title[Do Coverage and Mutation Scores of LLM-Generated Test Suites Correlate with Their Effectiveness?]{Do Coverage and Mutation Scores of LLM-Generated Test Suites Correlate with Their Effectiveness? (Replicability Study)}

\author{Junda Zhao}
\orcid{0000-0003-4978-4128}
\affiliation{%
  \institution{University of Toronto}
  \department{Department of Mechanical and Industrial Engineering}
  \city{Toronto}
  \country{Canada}
}
\email{junda.zhao@mail.utoronto.ca}

\author{Shurui Zhou}
\orcid{0000-0002-6346-6073}
\affiliation{%
  \institution{University of Toronto}
  \department{Department of Electrical and Computer Engineering}
  \city{Toronto}
  \country{Canada}
}
\email{shurui.zhou@utoronto.ca}

\author{Eldan Cohen}
\orcid{0000-0001-5767-6683}
\affiliation{%
  \institution{University of Toronto}
  \department{Department of Mechanical and Industrial Engineering}
  \city{Toronto}
  \country{Canada}
}
\email{eldan.cohen@utoronto.ca}

\renewcommand{\shortauthors}{Junda Zhao, Shurui Zhou, and Eldan Cohen}

\begin{abstract}
Recent advances in large language models (LLMs) have driven growing interest in using LLMs to automate test generation. Prior work commonly evaluates generated test suites using proxy metrics such as code coverage and mutation score. However, studies by Inozemtseva et al.\ and Papadakis et al.\ show that, for human-written tests, correlations among coverage, mutation, and real-bug detection can largely vanish once test suite size is controlled, raising concerns about the validity of evaluations based on proxy metrics. It also remains unclear whether these conclusions carry over to LLM-generated tests, given that prevailing LLM-based test-generation workflows differ substantially from traditional approaches.

In this paper, we conduct a large-scale replication study of these two prior works using a wide range of test suites generated by a diverse set of LLMs, and re-examine the relationships among coverage, mutation, and real-bug detection effectiveness. Our findings diverge substantially from prior results. We show that the usefulness of coverage and mutation is highly context-dependent: in regression-style settings where the code provided to the LLM can be reasonably assumed bug-free, these metrics can provide meaningful signals when comparing \emph{across} models; in another common scenario where the code-under-test may already be buggy and the goal is to expose the bug within the code-under-test, they no longer serve as reliable indicators. We also find little evidence that test suite size is a dominant confounder for correlations among coverage, mutation, and real-bug detection for LLM-generated tests. Based on these findings, we discuss how to interpret results from prior studies and provide actionable guidance for evaluating LLM-based test generation.

\end{abstract}




\maketitle

\section{Introduction}
\label{sec:Introduction}

Software testing plays an indispensable role in software quality assurance, yet manually creating and maintaining extensive test suites remains resource-intensive and error-prone, motivating continued interest in automating this process. Recent advances in large language models (LLMs) have sparked growing interest in using these models to automatically generate test suites~\cite{ChatUniTestFSE2024Demo, EvalImprChatgptUTGenFSE2024, TestGenLLMandMutIST, OnLLMUTGenEvalASE2024, EmpEvalforLLMUTAutoGenTSE}. In this line of work, the effectiveness of LLM-generated test suites is most commonly assessed using code coverage metrics (e.g., statement, branch, and modified condition coverage), alongside basic statistics such as the number of generated tests, compilation success rate, and pass rate~\cite{ChatUniTestFSE2024Demo, EvalImprChatgptUTGenFSE2024, TestGenLLMandMutIST, OnLLMUTGenEvalASE2024, EmpEvalforLLMUTAutoGenTSE}. While these metrics provide useful signals about test thoroughness, they do not directly measure whether generated tests can expose real bugs~\cite{unittestgenerationreview}. To move beyond coverage, some studies adopt mutation score as a proxy for test effectiveness, leveraging its ability to approximate fault detection without requiring manually curated real-world bugs~\cite{TestGenLLMandMutIST, zhang2024testbenchevaluatingclassleveltest, abdullin2025testwarscomparativestudy}. However, despite mutation testing being a well-established technique, its validity as a surrogate for real-bug detection remains debated as mutation operators may not reflect the characteristics or complexity of real bugs encountered in practice~\cite{papadakis_mutation_vs_real_bug_detect}. Overall, direct evaluation of the core objective of testing---namely, the real-bug detection effectiveness of generated test suites---remains comparatively limited in the LLM-based test generation literature.

This naturally raises the question of whether these widely used proxy metrics are actually indicative of the \emph{bug detection effectiveness} of LLM-generated test suites. The testing literature has extensively examined the relationships among coverage, mutation, and bug detection for human-written tests, but has reached mixed conclusions~\cite{corr_cov_bug_human_4, corr_cov_bug_human_2, corr_cov_bug_human_3, corr_mut_bug_1, corr_mut_bug_2}. Notably, Inozemtseva et al.~\cite{CovNotStronglyCorrWithEffICSE2014} show that the correlation between coverage and test effectiveness (using mutation as a proxy) can largely disappear once test suite size is controlled, and Papadakis et al.~\cite{papadakis_mutation_vs_real_bug_detect} report analogous findings for the relationship between mutation score and real-bug detection effectiveness. These results raise a key concern for current LLM-based evaluations: if coverage and mutation score do not reliably track bug detection effectiveness, then evaluating LLM-generated tests primarily through these proxies may be misleading. At the same time, whether conclusions drawn for human-written tests carry over to LLM-generated tests remains uncertain. LLM-based test generation typically uses the code-under-test as the primary prompt input~\cite{ChatUniTestFSE2024Demo, EvalImprChatgptUTGenFSE2024, TestGenLLMandMutIST, OnLLMUTGenEvalASE2024, EmpEvalforLLMUTAutoGenTSE}, yielding a predominantly white-box setting that differs from that of many human-written tests, which are often derived from program requirements rather than synthesized entirely from the implementation~\cite{testing_textbook}. Together, these factors call for a dedicated re-examination of how coverage and mutation relate to real-bug detection for LLM-generated test suites.

In this paper, we conduct a large-scale replication study of two prior studies by Inozemtseva et al.\ and Papadakis et al.\ to re-examine the relationships among test suite size, test coverage, mutation score, and real-bug detection effectiveness for LLM-generated test suites across 11 state-of-the-art (SOTA) LLMs. Our study focuses on LLM-generated tests, whereas prior studies primarily studied human-written tests, with Papadakis et al.\ also including tests generated by EvoSuite. Moreover, LLM-generated tests are commonly produced in a predominantly white-box setting, where the code-under-test is provided as the primary prompt input. This input may be either bug-free or buggy, which can substantially affect the behavior of the generated tests~\cite{huang2025measuringinfluenceincorrectcode} and motivates further investigation of how this factor influences the correlations studied in prior work. Given these differences, we frame our study as a conceptual replication~\cite{Concept_rep_def_1, Concept_rep_def_2} rather than a direct replication. Specifically, we revisit a similar set of research questions by examining whether the previously observed correlations among coverage, mutation, and real-bug detection effectiveness continue to hold for LLM-generated tests using both bug-free and buggy code as input, while adapting the research procedures to this new setting. We describe these methodological adaptations in Section~\ref{sec:Methodology}.

Our results for LLM-generated tests diverge substantially from the conclusions reported in both prior studies; we summarize the main takeaways here. Although Inozemtseva et al.\ and Papadakis et al.\ question whether coverage and mutation metrics reliably indicate bug detection effectiveness, we find that these metrics can be informative when the code-under-test can be reasonably assumed bug-free, and the goal is regression testing (i.e., to catch bugs introduced by future code changes), even when test suite size is controlled. In this setting, certain coverage metrics and mutation scores provide useful signals when comparing \emph{across} models. However, in the more general and practically challenging setting where the correctness of the code-under-test provided to the LLM cannot be guaranteed, and the generated tests are expected to detect bugs present in that code, coverage becomes unreliable as an indicator of bug detection effectiveness, and mutation analysis is not applicable. Furthermore, unlike both prior studies, we find little evidence that the size of LLM-generated test suites (i.e., the number of tests) is a strong confounding factor in the relationships among coverage, mutation score, and real-bug detection effectiveness.

In summary, this paper makes the following contributions:
\newcommand{\myitem}[1]{%
\par\noindent
\hangindent=10pt
\hangafter=1
\makebox[10pt][l]{\labelitemi}#1\par
}

\myitem{We replicate the studies of Inozemtseva et al.\ and Papadakis et al.\ on the relationships among coverage, mutation, and bug detection effectiveness using over 100{,}000 test cases generated by 11 SOTA LLMs, and show that the resulting correlations for LLM-generated tests differ substantially from those reported in the original studies on human-written tests.}

\myitem{We extend Inozemtseva et al.'s study beyond mutants to real-world bugs by examining how coverage correlates with real-bug detection for test suites generated from both bug-free and buggy code. This allows us to characterize how the coverage--effectiveness relationship changes in practically relevant settings where the code provided to the LLM may itself be buggy.}

\myitem{Based on our findings, we discuss how to interpret coverage and mutation metrics in the LLM era and provide actionable guidance for using these metrics and for evaluating the quality of LLM-generated tests in future work.}

\myitem{Following best practices for replication studies, we provide an artifact containing our implementation, generated data, and analysis scripts to support future replication and secondary studies on coverage, mutation, and bug detection effectiveness for LLM-based unit test generation.}

\begin{table*}[htb]
\vspace{-0.6em}
\centering
\tiny
\caption{Studies focusing on evaluating LLMs for test generation.}
\vspace{-0.6em}
\label{tab:related-works}
\begin{tabularx}{\textwidth}{@{\extracolsep{\fill}}
    >{\centering\arraybackslash}p{0.7cm}
    p{1.5cm}   
    p{3.1cm}   
    p{3.1cm}   
    p{2.6cm}   
    c
}

\toprule
\textbf{Citation} & \textbf{Models} & \textbf{Datasets} & \textbf{Metrics} & \textbf{Prompt Content} & \textbf{Buggy Code}\\
\midrule

\cite{UnitTestCaseGenTrans} & BART & METHOD2TEST \cite{UnitTestCaseGenTrans}, Defects4J & Correctness, Coverage &  Code-under-test, Code context & \ding{55} \\
\midrule
\cite{EmpEvalforLLMUTAutoGenTSE} & GPT-3.5 & Public code repo, \newline Defects4J & Correctness, Coverage & Code-under-test, Code context, \newline Example usage & \ding{55} \\
\midrule
\cite{InitialInvestigationChatGPT} & GPT-3.5-turbo & Public code repo & Correctness, Coverage, Mutation & Code-under-test & \ding{55}\\
\midrule
\cite{CAT-LMASE2023} & GPT-2 & Public code repo & Correctness, Coverage, \newline Lexical similarity & Code-under-test & \ding{55} \\
\midrule
\cite{ChatVSSBSTTSE}& GPT-3 & DynaMOSA \cite{dynamosa}, Defects4J  & Correctness, Coverage, \newline Bug detection, User study & Code-under-test & \ding{55} \\
\midrule
\cite{LLMforJunitTestGenEASE2024} & Codex, \mbox{StarCoder}, \newline GPT-3.5-turbo & HumanEval \cite{HumanEval}, SF110 \cite{Evosuite} & Correctness, Coverage, Test style & Code-under-test, Code context & \ding{55}\\
\midrule
\cite{EvalImprChatgptUTGenFSE2024} & GPT-3.5-turbo & CodeSearchNet & Correctness, Coverage, User study & Code-under-test, \newline LLM-generated code intention & \ding{55}\\
\midrule
\cite{UTGenUsingGenAIICSEWorkShop2024} & GPT-3.5 & Public code repo & Correctness, Coverage & Code-under-test & \ding{55}\\
\midrule
\cite{OnLLMUTGenEvalASE2024} & 6 models & Defects4J & Correctness,  Coverage, \newline Bug detection & Code-under-test, Code context & \ding{55} \\
\midrule
\cite{zhang2024testbenchevaluatingclassleveltest} & CodeLlama, \newline GPT-3.5, GPT-4 & Public code repo  & Correctness, Coverage, Mutation & Code-under-test, Code context & \ding{55} \\
\midrule
\cite{sysforautocreateandassess} & GPT-3.5-turbo, \newline GPT-4 & Classes2Test \cite{sysforautocreateandassess} & Correctness, Coverage, Mutation, \newline  Test style & Code-under-test, Code context  & \ding{55} \\
\midrule
\cite{LargeScaleEmpStudyFTUTGenISSTA2025} & 37 models & METHOD2TEST, Defects4J, \newline ATLAS \cite{ATLAS}, Ceprot \cite{Ceprot}  & Correctness, Lexical similarity  & Code-under-test & \ding{55} \\
\midrule
\cite{jain2025testgeneval} & 10 models & TestGenEval \cite{jain2025testgeneval} & Correctness, Coverage, Mutation & Code-under-test, Code context & \ding{55} \\
\midrule
\cite{wang-etal-2025-testeval} & 17 models & TestEval \cite{wang-etal-2025-testeval} & Correctness, Coverage & Code-under-test & \ding{55}\\
\midrule
\cite{abdullin2025testwarscomparativestudy} & 4 models & GitBug-Java \cite{Gitbug} & Correctness, Coverage, \newline Mutation, Bug detection &  Code-under-test, Code context & \ding{55} \\
\midrule
\cite{huang2025measuringinfluenceincorrectcode} & 11 models & HumanEval \cite{HumanEval}, MBPP \cite{mbpp}, APPS \cite{apps}, SWE-Bench \cite{jimenez2024swebench}, BugsInPy \cite{BugsInPy}  & Correctness, Coverage, \newline Bug detection &  Code-under-test, Code context, \newline Code documentation & \ding{51} \\

\bottomrule
\end{tabularx}
\vspace{-1.5em}
\end{table*}

\section{Related Works}
In this section, we contextualize the motivation in Section~\ref{sec:Introduction} by reviewing recent work on evaluating LLM-based unit test generation. We focus on studies that assess LLMs' test-generation capability in broadly applicable settings, as these works tend to yield more generalizable conclusions. We exclude methodology papers that propose specific techniques to improve LLM-based test generation, since they typically assume a predefined usage scenario and success criteria (e.g., maximizing coverage) and often evaluate in a narrower setting, making their conclusions less generalizable than those of general evaluation studies. We compile the reviewed studies based on the latest survey on LLM-based test generation~\cite{unittestgenerationreview} and summarize them in Table~\ref{tab:related-works}.

Across the reviewed studies, correctness metrics (statistics of tests that compile or pass) and code coverage are by far the most commonly reported outcomes. Five studies report mutation results, and four evaluate real-bug detection. We also observe that all studies use the code-under-test as the primary prompt input, consistent with a predominantly white-box test-generation paradigm. Finally, except for Huang et al.~\cite{huang2025measuringinfluenceincorrectcode}, no prior study considers the setting in which the input code itself may be buggy. Together, these observations motivate our investigation of whether commonly used evaluation metrics (e.g., correctness, coverage, and mutation) are indicative of bug detection effectiveness, and whether they remain informative when tests are generated from buggy inputs.

A natural question is why these studies converge on a similar set of evaluation metrics. Our review suggests that this evaluation setup can be traced to the earliest transformer-based unit test generation work~\cite{UnitTestCaseGenTrans}. Tracing further back, we also observe that the same metric choices were commonly adopted in the pre-LLM era for evaluating automated test generation tools such as EvoSuite~\cite{whitebox_test, automatedunittestevosuiteregress}. These tools are largely white-box and generate regression tests that encode the behavior of the program version given as input, typically assuming that the input code is bug-free. Under this assumption, evaluations are largely limited to proxy metrics such as coverage and mutation, since directly measuring real-bug detection on the code-under-test is not meaningful when that code is presumed correct.

In the LLM era, however, test generation is not inherently limited to this regression-testing paradigm. Unlike prior tools that mainly infer tests from the implementation, LLMs can, in certain cases, infer intended behavior beyond the code itself. For example, Huang et al.~\cite{huang2025measuringinfluenceincorrectcode} show that LLMs can still generate bug-detecting tests from buggy code, although less frequently than from bug-free code. Hossain et al.~\cite{Doc2OracLL} further show that additional context, such as documentation, can improve LLMs' ability to infer intended behavior from potentially buggy inputs. Together, these findings suggest that LLM-based test generation need not rely solely on the code-under-test or on the assumption that the implementation is correct. Nevertheless, most existing evaluations of LLM-based test generation largely inherit these evaluation settings from pre-LLM tools by default and, to the best of our knowledge, few studies explicitly examine what these metrics actually reflect in the LLM setting. We therefore argue that analyzing these widely adopted metrics for LLM-generated tests is necessary to interpret prior results correctly and to clarify what conclusions can (and cannot) be drawn from them.

\section{Methodology}
\label{sec:Methodology}

\subsection{Research Questions}
We structure our study around the following research questions:
{\emergencystretch=2em
\begin{itemize}[leftmargin=10pt]
  \item \textbf{RQ1: Is test coverage correlated with mutation score for LLM-generated test suites?}
  We replicate Inozemtseva et al.'s correlation analyses in the LLM setting, examining the relationship between test suite size and mutation score, coverage--mutation correlations under both unconstrained and size-controlled suites, and correlations among different coverage criteria.

  \item \textbf{RQ2: Is test coverage correlated with real-bug detection effectiveness for LLM-generated test suites?}
  We move beyond mutation to real bugs, assessing whether the correlations observed in RQ1 persist when effectiveness is measured via real-bug detection. We further consider the practically relevant setting in which the code provided to the LLM is buggy, and analyze coverage--bug detection correlations for tests generated from both bug-free and buggy code.\footnote{We do not conduct mutation testing, or analyze the correlation between mutation score and bug detection, for buggy code in RQ2 because mutation testing presupposes a passing (``green'') test suite on the code-under-test; this would exclude precisely the tests that expose the bug within the code-under-test, hiding the very bug to be detected and rendering the resulting mutation scores meaningless. Mutation analysis is therefore not applicable in this setting.}

  \item \textbf{RQ3: Is mutation score correlated with real-bug detection effectiveness for LLM-generated test suites?}
  Using the data collected for RQ1 and RQ2, we replicate Papadakis et al.'s mutation--bug detection correlation analysis to evaluate whether the mutation score is predictive of real-bug detection for LLM-generated test suites.
\end{itemize}}

\subsection{Benchmark}

In this study, we focus on large Java projects, following the scope of Inozemtseva et al. Specifically, we use Defects4J~\cite{Defects4j}, a widely adopted benchmark of real bugs from open-source Java projects (e.g., JFreeChart and Apache Commons Lang) and a common choice in recent evaluations of automated and LLM-based unit test generation~\cite{UnitTestCaseGenTrans, A3TestIST, EmpEvalforLLMUTAutoGenTSE, OnLLMUTGenEvalASE2024}. Defects4J was also used by Papadakis et al.\ to study the relationship between mutation score and real-fault detection effectiveness. Each defect includes both a buggy version and its corresponding fixed (bug-free) version, along with developer-written tests that serve as test oracles. Defects4J v3.0 contains 854 defects across 17 projects, and each defect may modify one or more methods; we refer to these patched methods as \emph{focal methods}. These focal methods constitute the units under test and form the core code context that we provide to LLMs for unit test generation.

Following prior work on evaluating LLM-based unit test generation \cite{OnLLMUTGenEvalASE2024, EvalImprChatgptUTGenFSE2024, EmpEvalforLLMUTAutoGenTSE}, we focus test generation and evaluation on focal methods. To select focal methods, we follow the general practice of prior studies \cite{OnLLMUTGenEvalASE2024, A3TestIST, UnitTestCaseGenTrans} and further refine the selection based on the following criteria:
\begin{enumerate}[leftmargin=*]
    \item Instead of retaining only public methods, we keep all \emph{non-private} focal methods, since public, protected, and package-private methods can be accessed by tests placed in the same package.
    \item Each focal method must exist in both the buggy and fixed versions to enable meaningful comparisons; we exclude cases where a focal method was removed or added after patching.
    \item We retain only focal methods whose buggy version triggers at least one developer-written failing test in Defects4J, ensuring that the corresponding patch reflects a functional bug fix rather than a refactoring or performance-only change.
\end{enumerate}

In total, our curated benchmark comprises 318 buggy focal methods spanning all 17 projects in Defects4J.

\subsection{Models}
\label{subsec:models}
To support a comprehensive and robust analysis, we evaluate 11 state-of-the-art (SOTA) large language models from six major developers, covering both \emph{base} and \emph{reasoning} variants where available. Reasoning models are trained to explicitly perform multi-step reasoning before producing an answer, whereas base models generate outputs directly. For hybrid models that can operate in either mode (e.g., Claude~4~Sonnet~\cite{anthropic2025claude4sonnet}), we evaluate both configurations, with reasoning enabled and disabled. Overall, this yields 13 model settings; for each setting, we generate test suites for both the buggy and fixed versions of every focal method. This breadth of models and configurations reduces the likelihood that our findings are driven by a single model family or a particular generation style, and enables a more reliable examination of how coverage and mutation score relate to real-bug detection for LLM-generated unit tests. Table~\ref{tab:evaluated_models} presents the list of all evaluated models.

\begin{table}[htbp!]
\vspace{-0.6em}
\centering
\scriptsize
\caption{LLMs used in our study.}
\vspace{-0.6em}
\label{tab:evaluated_models}
\begin{tabular}{lllc}
\toprule
\textbf{Developer} & \textbf{Model Name} & \textbf{Type} & \textbf{Open Sourced} \\
\midrule
\multirow{2}{*}{Google} & \emph{Gemini 2.5 Pro} \cite{Google2025Gemini2.5Pro} & Reasoning & \ding{55} \\
 & \emph{Gemini 2.5 Flash} \cite{Google2025Gemini2.5Flash} & Hybrid & \ding{55} \\
\cmidrule(lr){1-4}
Anthropic & \emph{Claude 4 Sonnet} \cite{anthropic2025claude4sonnet} & Hybrid & \ding{55} \\
\cmidrule(lr){1-4}
\multirow{2}{*}{xAI} & \emph{Grok-4} \cite{xai2025grok4} & Reasoning & \ding{55} \\
 & \emph{Grok-3} \cite{xai2025grok3} & Base & \ding{55} \\
\cmidrule(lr){1-4}
\multirow{2}{*}{OpenAI} & \emph{GPT-4.1} \cite{openai2025gpt41} & Base & \ding{55} \\
 & \emph{GPT-O4-mini} \cite{openai2025o4mini} & Reasoning & \ding{55} \\
\cmidrule(lr){1-4}
\multirow{2}{*}{DeepSeek} & \emph{DeepSeek-V3} \cite{deepseekai2025deepseekv3technicalreport} & Base & \ding{51} \\
 & \emph{DeepSeek-R1} \cite{deepseekai2025deepseekr1incentivizingreasoningcapability} & Reasoning & \ding{51} \\
\cmidrule(lr){1-4}
\multirow{2}{*}{Alibaba} & \emph{Qwen3-Coder-Plus} \cite{alibaba2025qwen3coder} & Base & \ding{51} \\
 & \emph{Qwen3-Plus} \cite{alibaba2025qwen3} & Reasoning & \ding{51} \\
\bottomrule
\end{tabular}
\vspace{-1.5em}
\end{table}

\subsection{Test Generation Workflow}
\label{subsec:Test_gen_workflow}
To ensure a realistic and generalizable evaluation, we adopt a test-generation workflow that closely follows the comprehensive pipeline for evaluating LLM-generated unit tests proposed by Yang et al.~\cite{OnLLMUTGenEvalASE2024}, covering both prompt construction and unit test extraction. Guided by their findings on effective prompt context, we include additional, readily obtainable code features surrounding the focal method, including its parameters, the enclosing class constructor, declared fields, and other methods, as well as constructors of user-defined classes that appear as parameter types or as the return type. The focal method body is the central component of the prompt. We wrap this context with a system instruction that frames the LLM as a professional Java developer and append an explicit request to generate unit tests for the provided code. Due to space constraints, we omit the full prompt template here; it is available in our replication package.

We adopt Yang et al.'s prompt configuration for three reasons. First, Yang et al.\ evaluated a broad range of open- and closed-source models, supporting the generality of their prompt design. Second, they provide an extensive ablation study on the impact of additional code context, demonstrating the utility of the specific features they include. Third, the included context can be extracted solely from the code-under-test without manual inspection, enabling a fully automated prompt-construction and test-generation pipeline that aligns with the practical goal of reducing developer burden through automation.

For unit test extraction from LLMs' outputs, we utilize an abstract syntax tree (AST) parser (Tree-sitter \cite{tree-sitter}) to extract the generated unit tests and related information, such as imports and helper functions. We then combine the extracted content with project-specific dependencies and common dependencies required by the JDK (e.g., \texttt{java.util}) and JUnit (e.g., \texttt{org.junit.Assert}), prepared in advance to avoid compilation issues due to missing dependencies. 

We apply this pipeline to each focal method for each LLM, resulting in 8{,}268 generated test suites comprising 101{,}123 individual test cases. 

\subsection{Test Suite Construction}
For test suite construction, we follow the random sampling strategy of Inozemtseva et al.\ and Papadakis et al.\ but adapt it to reflect a more realistic test suite construction process. In particular, rather than sampling individual tests from a project-wide pool, we sample at the focal-method level. We avoid project-level sampling for two reasons. First, tests are typically written to exercise different aspects of a specific focal method, so project-wide sampling can mix tests from unrelated suites and does not reflect how test suites are constructed and used in practice. Second, when such mixed suites are size-controlled to a small number of tests, they often cover none of the mutants relevant to a given method, yielding many suites with mutation score $0$ and injecting noise into the correlation analysis.

\begin{figure*}[ht]
    \vspace{-0.6em}
    \centering
    \includegraphics[width=0.9\linewidth]{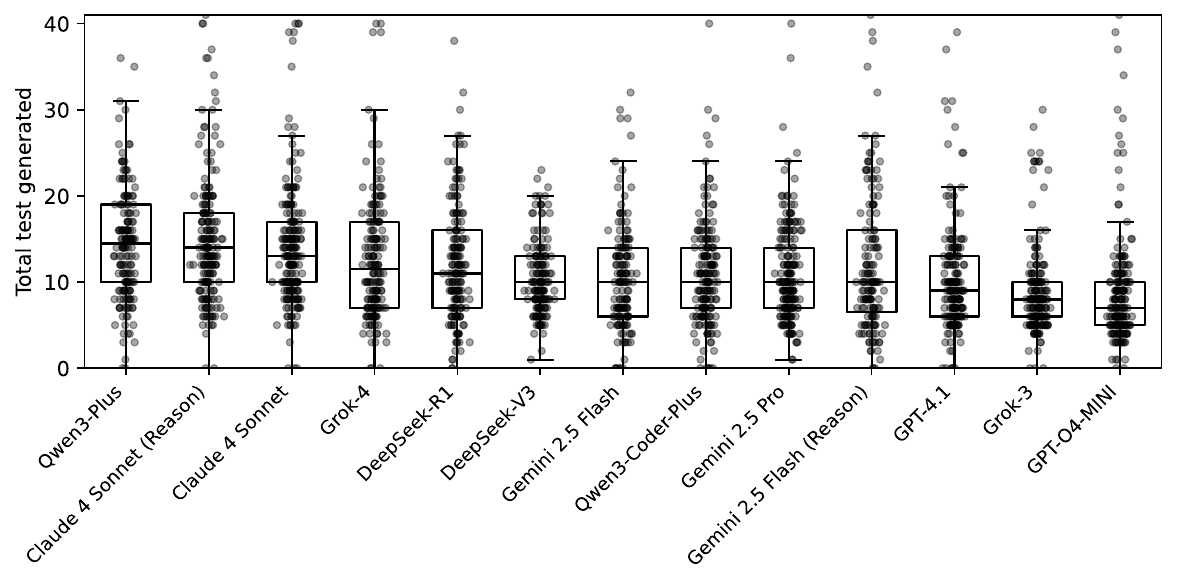}
    \vspace{-1.0em}
    \caption{Box plot of the number of tests within each test suite generated by different models.}
    \Description{Box plot showing the distribution of the number of tests per generated test suite for each of the thirteen evaluated LLM configurations.}
    \label{fig:total_test_number_boxplot}
    \vspace{-1.0em}
\end{figure*}

Concretely, for each draw we sample 100 focal methods, each paired with its corresponding LLM-generated test suite, from the set of generated suites that contain at least one compilable test. We construct 1{,}000 unique draws. In the unconstrained setting, we keep all tests in each selected suite, since LLMs naturally produce suites of varying sizes. In the size-controlled setting, because most models generate roughly 10 tests per suite on average (Figure~\ref{fig:total_test_number_boxplot}), we fix suite size by selecting $k \in \{3,5,10\}$ and retaining $k$ tests uniformly at random from each selected suite, restricting sampling to suites that contain at least $k$ tests. We do not consider smaller values of $k$ because very small suites frequently cover no mutants in the focal method, reintroducing the no-mutant-coverage issue noted above. We also avoid larger values of $k$ because too few suites contain at least $k$ tests, which would prevent us from constructing 1{,}000 unique draws.

\subsection{Measurements}
For the measurements required by our analysis, we largely follow the tools and metric definitions adopted by Inozemtseva et al.\ and Papadakis et al. We measure coverage with CodeCover~\cite{codecover}, collecting statement, branch, and modified condition coverage (MCC). We measure mutation with PIT~\cite{pit} and report both raw and normalized mutation scores using Inozemtseva et al.'s definitions, with a minor adaptation for our focal-method setting. Specifically, Inozemtseva et al.\ define raw mutation score as killed mutants divided by all mutants generated for the entire subject project, a ratio we instead compute over the mutants generated for the focal method associated with each test suite; their normalized mutation score, calculated as killed mutants divided by the mutants covered by the suite, is adopted unchanged. For real-bug detection, similar to Papadakis et al., we mark a buggy focal method as \emph{detected} if its generated test suite contains at least one \emph{effective} test that passes on the fixed version but fails on the buggy version, and we report bug detection effectiveness (the bug detection ratio) as the proportion of buggy focal methods detected.


Since our test suites are generated per focal method, we aggregate coverage and mutation in two complementary ways. In the \emph{average} aggregation, we compute coverage (or mutation score) per focal method with respect to that method and then take the mean across focal methods. In the \emph{accumulated} aggregation, we sum covered (or killed) elements and total elements over all focal methods and compute a single global ratio. Prior work effectively reports accumulated aggregates at the project level because its analyses are based on project-wide test pools. In our focal-method setting, accumulated values can be dominated by a small number of focal methods with many statements, conditions, or mutants, potentially obscuring per-method trends. We therefore report both aggregation schemes and compute correlations only within the same aggregation type (i.e., average with average, accumulated with accumulated). 

\subsection{Correlation Analysis}
To address our RQs, we perform three major correlation analyses: (i) coverage versus mutation score for generated test suites, (ii) coverage versus real-bug detection for test suites generated from both bug-free and buggy code, and (iii) mutation score versus real-bug detection.\footnote{We use the same Guilford scale \cite{guilfordscale} as Inozemtseva et al.\ for verbal descriptions: correlations with absolute value $<0.4$ are ``weak'' (low), $0.4$--$0.7$ are ``moderate'', $0.7$--$0.9$ are ``strong'', and $>0.9$ are ``very strong''.} As part of (i) and (ii), we also report correlations between test suite size and mutation score, and between test suite size and bug detection ratio---as in Inozemtseva et al.---to assess whether suite size confounds the relationships among the other metrics.


The large number of test suites generated by a diverse set of LLMs also enables us to examine correlations from complementary perspectives. For each correlation analysis, we report results at three granularities. (1) \emph{Combined}: we compute correlations over sampled draws of focal-method suites pooled across all models, providing an overall view that mixes within-model and between-model variation. (2) \emph{Intra-model}: we repeat the same analysis separately for each model to characterize how the relationship behaves among suites produced by a single LLM. (3) \emph{Inter-model}: instead of drawing random combinations of suites as in the first two views (and in prior work), we aggregate each model's generated test suite statistics (considering only compilable tests) into a single data point per model---for metrics such as suite size, coverage, mutation, and bug detection---and then compute correlations across models. Although this view has a smaller sample size, it can still reveal statistically meaningful patterns and help disentangle whether observed correlations are driven primarily by differences between models or by variation within a model.

\section{RQ1: Correlation Between Coverage and Mutation Score}
\label{sec:Study_Results}
For RQ1, we replicate the study of Inozemtseva et al.\ by analyzing the correlation between coverage and test effectiveness (using mutation score as a proxy) for LLM-generated tests. We mirror the structure of the original work and report results for the same set of correlation analyses.

\subsection{Is Size Correlated With Mutation Score?}
Inozemtseva et al.\ reported a strong positive correlation between test suite size and both raw and normalized mutation score. To examine whether this conclusion carries over to LLM-generated tests, we conduct a similar analysis with one key difference. Rather than constructing suites of pre-specified sizes by sampling from a project-wide test pool, as in prior work, we leverage the natural variation in LLM-generated test suite sizes. Our randomly sampled LLM-generated test suites already span a wide range of test counts, allowing us to compute correlations directly between the number of generated tests and mutation score. This setup also better reflects why suite sizes vary in practice, where differences are largely driven by the characteristics and testing needs of the code-under-test rather than by manually enforcing a target size.

\begin{table}[htbp!]
\vspace{-0.6em}
\centering
\caption{Correlation between the number of tests (averaged for average aggregation and summed for accumulated aggregation) and both types of mutation scores over test suites from all models; all entries are significant ($p<0.05$).}
\vspace{-0.6em}
\label{tab:test_size_vs_mutation}
\scriptsize
\setlength{\tabcolsep}{6pt}
\begin{tabular}{lcccc}
\toprule
& \multicolumn{2}{c}{Average aggregation} & \multicolumn{2}{c}{Accumulated aggregation} \\
\cmidrule(lr){2-3}\cmidrule(lr){4-5}
& Pearson $r$ & Kendall $\tau$ & Pearson $r$ & Kendall $\tau$ \\
\midrule
Raw mutation score
& 0.031 & 0.034 & 0.029 & 0.017 \\
Normalized mutation score
& 0.098 & 0.066 & -0.112 & -0.047 \\
\bottomrule
\end{tabular}
\end{table}

\begin{figure*}[ht]
    \centering
    \includegraphics[width=0.9\linewidth]{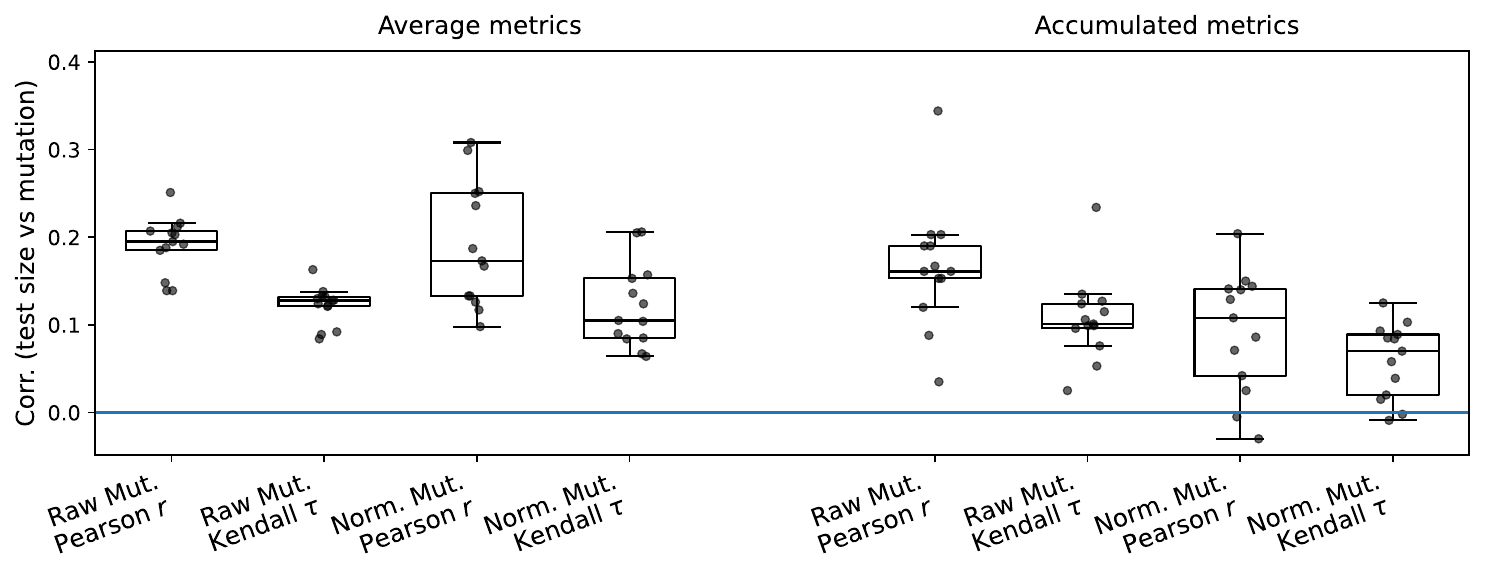}
    \vspace{-1.0em}
    \caption{Box plots of the correlation between test suite size and mutation score (raw and normalized) in the unconstrained setting. Each dot is the correlation computed from model-specific sampled test suite combinations; boxes group these correlations by aggregation type (average vs.\ accumulated) and statistic (Pearson $r$ vs.\ Kendall $\tau$).}
    \Description{Box plots summarizing per-model correlation coefficients between test suite size and raw and normalized mutation scores.}
    \label{fig:test_size_vs_mut_per_mode_boxplot}
    \vspace{-1.0em}
\end{figure*}

We begin with the combined view. As shown in Table~\ref{tab:test_size_vs_mutation}, the correlations between test suite size and both raw and normalized mutation scores are uniformly weak under both average and accumulated aggregation. We then examine the same relationship from the intra-model view to assess whether the trend varies across LLMs. Figure~\ref{fig:test_size_vs_mut_per_mode_boxplot} reports the within-model correlations; compared to the pooled analysis, these correlations are in many cases slightly higher but remain weak for every model. Finally, we consider the inter-model view: Table~\ref{tab:test_size_vs_mutation_aggregated_by_model} shows that these model-level correlations are also weak and not statistically significant. Overall, in contrast to Inozemtseva et al., we observe weak to negligible correlations between test suite size and both raw and normalized mutation score for LLM-generated test suites across combined, intra-model, and inter-model analyses. This suggests that LLMs that generate more tests do not necessarily produce test suites with stronger mutant-killing capability.

\begin{table}[htbp!]
\centering
\caption{Inter-model correlation between the number of tests (averaged for average aggregation and summed for accumulated aggregation) and mutation scores; $p$-values are reported in parentheses.}
\label{tab:test_size_vs_mutation_aggregated_by_model}
\scriptsize
\setlength{\tabcolsep}{6pt}
\begin{tabular}{lcccc}
\toprule
& \multicolumn{2}{c}{Average aggregation} & \multicolumn{2}{c}{Accumulated aggregation} \\
\cmidrule(lr){2-3}\cmidrule(lr){4-5}
& Pearson $r$ & Kendall $\tau$ & Pearson $r$ & Kendall $\tau$ \\
\midrule
Raw mutation score
& 0.185 (0.546) & 0.205 (0.367) & 0.259 (0.393) & 0.231 (0.306) \\
Normalized mutation score
& 0.093 (0.761) & 0.026 (0.952) & -0.347 (0.245) & -0.231 (0.306) \\
\bottomrule
\end{tabular}
\end{table}

\subsection{Is Coverage Correlated With Mutation Score When Size Is Ignored?}
\label{subsec:cov_corr_mut_size_ignored}

\begin{figure*}[ht]
    \centering
    \begin{subfigure}{0.88\linewidth}
        \centering
        \includegraphics[width=\linewidth]{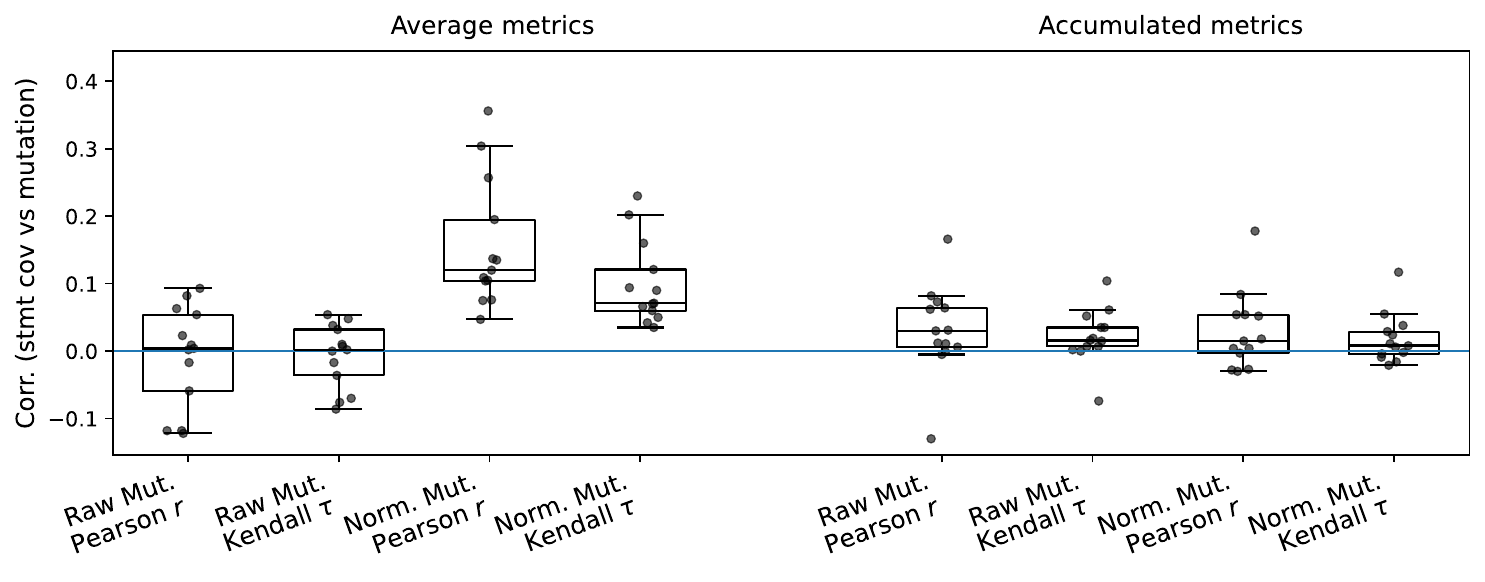}
        \vspace{-1.5em}
        \caption{Statement coverage vs.\ mutation score.}
        \label{fig:cov_mut_boxplot_stmt}

    \end{subfigure}

    \begin{subfigure}{0.88\linewidth}
        \centering
        \includegraphics[width=\linewidth]{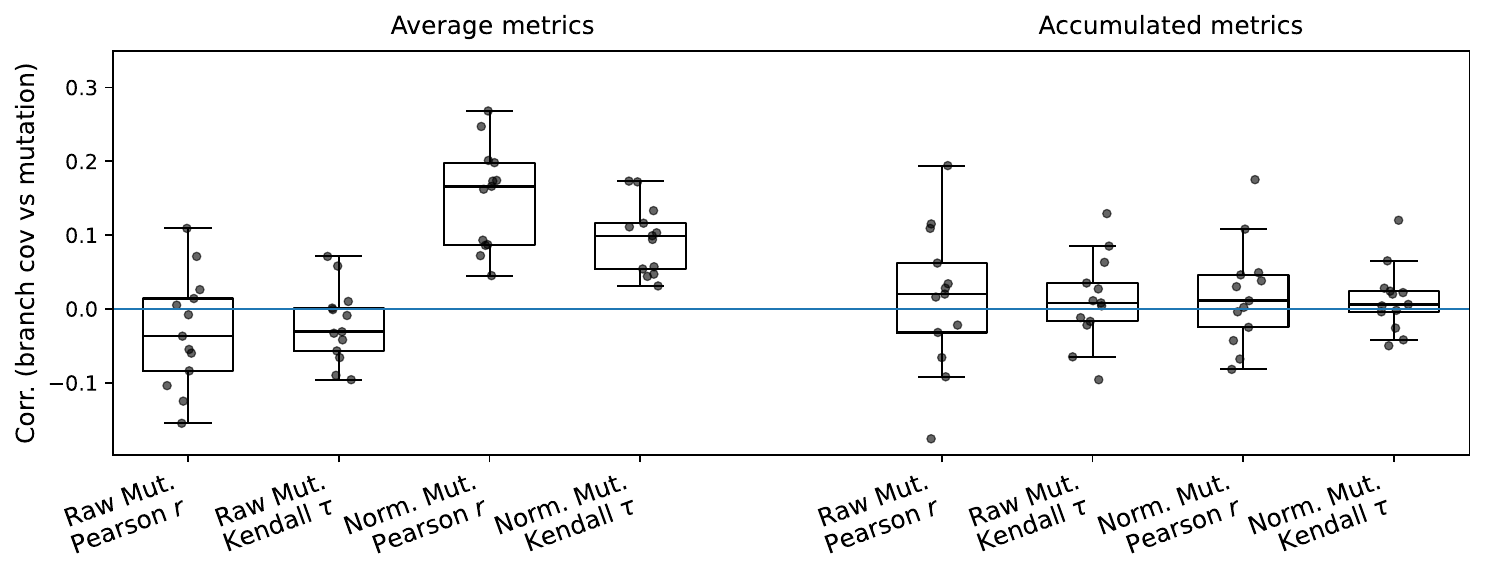}
        \vspace{-1.5em}
        \caption{Branch coverage vs.\ mutation score.}
        \label{fig:cov_mut_boxplot_branch}
    \end{subfigure}

    \begin{subfigure}{0.88\linewidth}
        \centering
        \includegraphics[width=\linewidth]{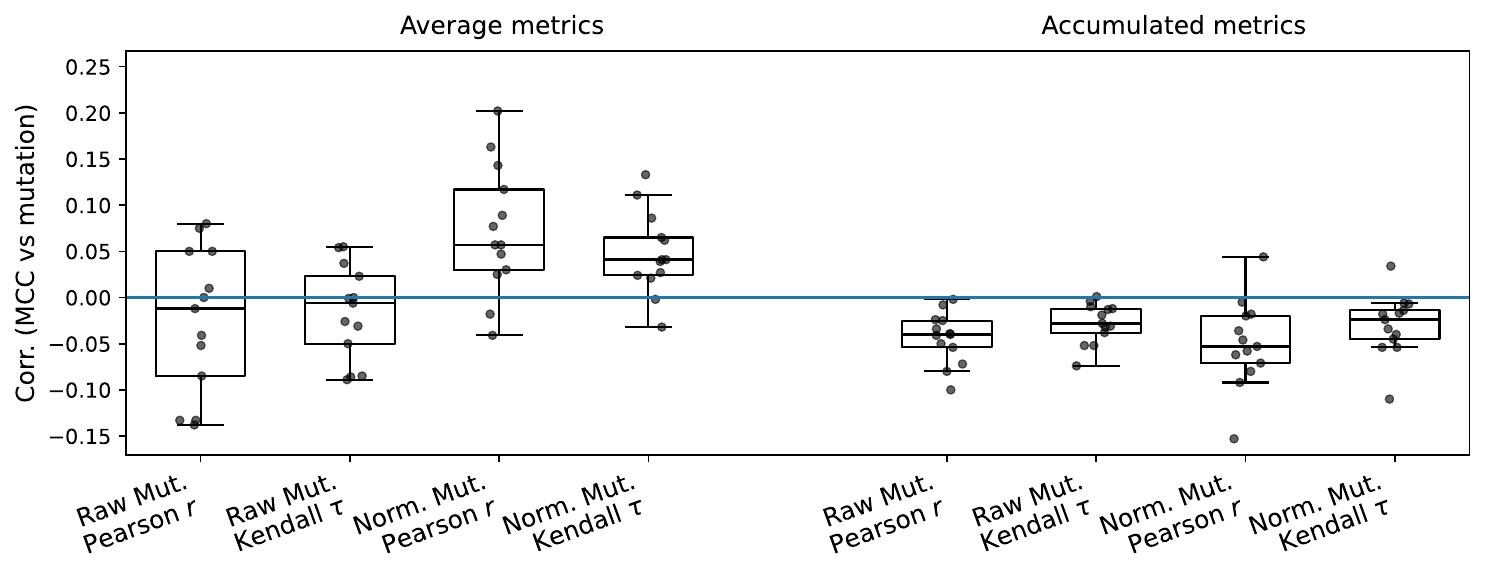}
        \vspace{-1.5em}
        \caption{Modified condition coverage vs.\ mutation score.}
        \label{fig:cov_mut_boxplot_mcc}
    \end{subfigure}

    \caption{Box plots of the correlation between coverage (statement, branch, and modified condition) and mutation score (raw and normalized) when test suite size is ignored. Each dot is the correlation computed from model-specific sampled test suite combinations; boxes group these correlations by aggregation type (average vs.\ accumulated) and statistic (Pearson $r$ vs.\ Kendall $\tau$).}
    \Description{Box plots summarizing per-model correlation coefficients between statement, branch, and modified condition coverage and raw and normalized mutation scores.}
    \label{fig:cov_vs_mut_per_model_boxplots}
    \vspace{-1.0em}
\end{figure*}

We next examine the relationship between coverage and mutation score without controlling for test suite size. Table~\ref{tab:cov_vs_mutation_all_models} summarizes the combined-view results. Correlations are consistently weak across coverage types and mutation-score variants under both average and accumulated aggregation: nearly all coefficients are below $0.4$, with the only exception being the Pearson correlation between average statement coverage and average normalized mutation score ($0.443$), which is just above the ``moderate'' threshold. The same pattern holds within individual models. As shown in Figure~\ref{fig:cov_vs_mut_per_model_boxplots}, intra-model correlations are typically even weaker than those in the pooled analysis: coefficients cluster tightly around zero and, in most settings, remain below $0.2$.

\begin{table*}[htbp!]
\centering
\caption{Correlation between coverage and mutation score across combinations of randomly selected test suites from all models, reported for both average aggregation and accumulated aggregation; all entries are significant ($p<0.05$), except the accumulated-aggregation correlation between normalized mutation score and statement coverage.}
\label{tab:cov_vs_mutation_all_models}
\scriptsize
\setlength{\tabcolsep}{6pt}
\begin{tabular}{llcccc}
\toprule
Coverage & Mutation score & \multicolumn{2}{c}{Average aggregation} & \multicolumn{2}{c}{Accumulated aggregation} \\
\cmidrule(lr){3-4}\cmidrule(lr){5-6}
& & Pearson $r$ & Kendall $\tau$ & Pearson $r$ & Kendall $\tau$ \\
\midrule
\multirow[c]{2}{*}{Statement}   & Raw        & 0.356 & 0.250 & 0.206 & 0.162 \\
                & Normalized & 0.443 & 0.272 & -0.009 & -0.004 \\
\midrule
\multirow[c]{2}{*}{Branch} & Raw        & 0.379 & 0.270 & 0.142 & 0.109 \\
                & Normalized & 0.348 & 0.238 & 0.032 & 0.022 \\
\midrule
\multirow[c]{2}{*}{MCC}  & Raw        & 0.267 & 0.186 & 0.118 & 0.092 \\
                & Normalized & 0.339 & 0.218 & -0.087 & -0.064 \\
\bottomrule
\end{tabular}
\end{table*}

The even weaker correlations under the intra-model view suggest that the slightly stronger correlations in the combined analysis might be driven by differences between models rather than by variation within a single model. We confirm this in the inter-model analysis. Table~\ref{tab:cov_vs_mutation_aggregated_per_model} shows that, under average aggregation, correlations increase noticeably for most coverage--mutation combinations. Although this view has a small sample size (13 data points), most coefficients remain statistically significant ($p<0.05$), indicating that models achieving higher average coverage also tend to achieve higher average mutation scores.


\begin{table*}[htbp!]
\centering
\caption{Correlation between coverage and mutation score with test statistics aggregated into one data point for each model, reported for both average aggregation and accumulated aggregation when test suite size is ignored; $p$-values are reported in parentheses.}
\label{tab:cov_vs_mutation_aggregated_per_model}
\scriptsize
\setlength{\tabcolsep}{6pt}
\begin{tabular}{llcccc}
\toprule
Coverage & Mutation score & \multicolumn{2}{c}{Average aggregation} & \multicolumn{2}{c}{Accumulated aggregation} \\
\cmidrule(lr){3-4}\cmidrule(lr){5-6}
& & Pearson $r$ & Kendall $\tau$ & Pearson $r$ & Kendall $\tau$ \\
\midrule
\multirow[c]{2}{*}{Statement} & Raw
& 0.676 (0.011) & 0.462 (0.030)
& 0.383 (0.197) & 0.410 (0.057) \\
                & Normalized
& 0.738 (0.004) & 0.538 (0.010)
& -0.047 (0.880) & -0.051 (0.858) \\
\midrule
\multirow[c]{2}{*}{Branch} & Raw
& 0.821 (0.001) & 0.641 (0.002)
& 0.246 (0.419) & 0.154 (0.510) \\
               & Normalized
& 0.596 (0.032) & 0.410 (0.057)
& 0.035 (0.909) & 0.051 (0.858) \\
\midrule
\multirow[c]{2}{*}{MCC}  & Raw
& 0.541 (0.057) & 0.385 (0.076)
& 0.289 (0.339) & 0.308 (0.163) \\
               & Normalized
& 0.625 (0.022) & 0.462 (0.030)
& -0.124 (0.686) & -0.051 (0.858) \\
\bottomrule
\end{tabular}
\end{table*}

\subsection{Is Coverage Correlated With Mutation Score When Size Is Fixed?}
We now analyze the correlation between coverage and mutation score after controlling for test suite size. Due to space constraints, we omit the full set of coefficients for the combined and intra-model views and summarize the results here. Overall, these size-controlled results mirror the unconstrained setting: whether we pool suites across all models or analyze each model separately, the correlation between coverage and mutation score remains weak across all tested suite sizes. The more interesting question is whether the moderate-to-strong inter-model correlation observed earlier persists once suite size is fixed.



\begin{table*}[htbp!]
\centering
\vspace{-1.0em}
\caption{Correlation between coverage and mutation score with test statistics aggregated into one data point for each model, for different fixed test suite sizes; $p$-values are reported in parentheses.}
\vspace{-0.6em}
\label{tab:cov_vs_mutation_aggregated_per_model_fixed_test_size}
\scriptsize
\setlength{\tabcolsep}{6pt}
\begin{tabular}{l l l cccc}
\toprule
Suite size $k$ & Coverage & Mutation score & \multicolumn{2}{c}{Average aggregation} & \multicolumn{2}{c}{Accumulated aggregation} \\
\cmidrule(lr){4-5}\cmidrule(lr){6-7}
& & & Pearson $r$ & Kendall $\tau$ & Pearson $r$ & Kendall $\tau$ \\
\midrule


\multirow[c]{6}{*}{3}
& \multirow[c]{2}{*}{Statement} & Raw
& 0.190 (0.534) & 0.103 (0.675)
& -0.144 (0.639) & -0.103 (0.675) \\
&                               & Normalized
& 0.642 (0.018) & 0.410 (0.057)
& -0.251 (0.409) & -0.128 (0.590) \\
& \multirow[c]{2}{*}{Branch}    & Raw
& 0.525 (0.066) & 0.282 (0.204)
& -0.102 (0.739) & -0.026 (0.952) \\
&                               & Normalized
& 0.579 (0.038) & 0.487 (0.022)
& 0.036 (0.907) & 0.000 (1.000) \\
& \multirow[c]{2}{*}{MCC}     & Raw
& 0.035 (0.910) & 0.000 (1.000)
& 0.156 (0.610) & 0.051 (0.858) \\
&                               & Normalized
& 0.669 (0.012) & 0.462 (0.030)
& 0.127 (0.680) & 0.128 (0.590) \\

\midrule

\multirow[c]{6}{*}{5}
& \multirow[c]{2}{*}{Statement} & Raw
& 0.476 (0.100) & 0.385 (0.076)
& 0.489 (0.090) & 0.282 (0.204) \\
&                               & Normalized
& 0.462 (0.112) & 0.282 (0.204)
& 0.009 (0.977) & 0.103 (0.675) \\
& \multirow[c]{2}{*}{Branch}    & Raw
& 0.699 (0.008) & 0.513 (0.015)
& -0.001 (0.997) & -0.154 (0.510) \\
&                               & Normalized
& 0.323 (0.282) & 0.256 (0.252)
& 0.165 (0.589) & 0.179 (0.435) \\
& \multirow[c]{2}{*}{MCC}     & Raw
& 0.476 (0.100) & 0.410 (0.057)
& 0.015 (0.962) & -0.077 (0.765) \\
&                               & Normalized
& 0.526 (0.065) & 0.410 (0.057)
& -0.283 (0.350) & -0.154 (0.510) \\

\midrule

\multirow[c]{6}{*}{10}
& \multirow[c]{2}{*}{Statement} & Raw
& 0.632 (0.021) & 0.590 (0.004)
& 0.140 (0.648) & 0.179 (0.435) \\
&                               & Normalized
& 0.688 (0.009) & 0.513 (0.015)
& -0.263 (0.386) & -0.282 (0.204) \\
& \multirow[c]{2}{*}{Branch}    & Raw
& 0.780 (0.002) & 0.692 (0.001)
& 0.054 (0.860) & -0.154 (0.510) \\
&                               & Normalized
& 0.528 (0.064) & 0.410 (0.057)
& 0.072 (0.816) & -0.103 (0.675) \\
& \multirow[c]{2}{*}{MCC}     & Raw
& 0.550 (0.052) & 0.462 (0.030)
& 0.095 (0.759) & 0.077 (0.765) \\
&                               & Normalized
& 0.652 (0.016) & 0.487 (0.022)
& -0.492 (0.088) & -0.333 (0.129) \\

\bottomrule
\end{tabular}
\vspace{-1.0em}
\end{table*}

Table~\ref{tab:cov_vs_mutation_aggregated_per_model_fixed_test_size} reports the inter-model results. Among coefficients that are statistically significant ($p<0.05$), the correlations generally decrease relative to the size-unconstrained setting, but they are typically moderate and, in some cases, strong (e.g., branch coverage vs.\ raw mutation score at $k{=}10$, $r{=}0.780$). This pattern differs from Inozemtseva et al., who reported that once suite size is controlled, coverage correlates weakly with normalized mutation score and only moderately with raw mutation score. Overall, our results suggest that even after controlling suite size, coverage can still provide some indication of mutant-killing strength at the inter-model level.

\subsection{Correlation Between Different Coverage Types}
Beyond analyzing coverage versus mutation score, Inozemtseva et al.\ also examined correlations among different coverage metrics and found very strong relationships across coverage types. We replicate this analysis for LLM-generated test suites. Table~\ref{tab:cov_cov_all_sizes} reports correlations among statement, branch, and modified condition coverage under both average and accumulated aggregation, for both unconstrained suites (All) and size-controlled suites.

Our results differ from those reported by Inozemtseva et al.\ in several respects. When suite size is unconstrained, coverage metrics are strongly correlated under average aggregation, but less so than the near-perfect relationships reported in 2014 (e.g., coefficients above 0.9 for both Pearson and Kendall $\tau$ between all coverage types). Pearson correlations typically remain strong, whereas Kendall $\tau$ is often only moderate. Under accumulated aggregation, statement and branch coverage remain strongly correlated, but correlations involving modified condition coverage often drop to moderate or weak. These gaps widen once suite size is controlled: under accumulated aggregation, modified condition coverage is sometimes only weakly correlated with statement and branch coverage, whereas under average aggregation the relationships remain mostly moderate to strong. Overall, these results do not support the claim from Inozemtseva et al.\ that different coverage types are essentially measuring the same thing in our LLM-generated setting.


\begin{table*}[htbp!]
\centering
\caption{Correlation between different coverage metrics (Statement, Branch, and Modified Condition) on sampled combinations of test suites, reported for the unconstrained setting (All) and for fixed suite sizes $k \in \{3,5,10\}$. $p$-values are reported in parentheses.}
\label{tab:cov_cov_all_sizes}
\scriptsize
\setlength{\tabcolsep}{6pt}
\begin{tabular}{l l cccc}
\toprule
Suite size $k$ & Coverage pair & \multicolumn{2}{c}{Average aggregation} & \multicolumn{2}{c}{Accumulated aggregation} \\
\cmidrule(lr){3-4}\cmidrule(lr){5-6}
& & Pearson $r$ & Kendall $\tau$ & Pearson $r$ & Kendall $\tau$ \\
\midrule

\multirow[c]{3}{*}{All}
& Statement vs Branch & 0.753 \,(0.0) & 0.553 \,(0.0) & 0.857 \,(0.0) & 0.675 \,(0.0) \\
& Statement vs MCC  & 0.885 \,(0.0) & 0.694 \,(0.0) & 0.319 \,($2.58\times 10^{-304}$) & 0.222 \,(0.0) \\
& Branch vs MCC    & 0.709 \,(0.0) & 0.510 \,(0.0) & 0.400 \,(0.0) & 0.271 \,(0.0) \\
\midrule


\multirow[c]{3}{*}{3}
& Statement vs Branch & 0.663 \,(0.0) & 0.468 \,(0.0) & 0.811 \,(0.0) & 0.614 \,(0.0) \\
& Statement vs MCC  & 0.789 \,(0.0) & 0.572 \,(0.0) & -0.003 \,(0.735) & 0.005 \,(0.408) \\
& Branch vs MCC     & 0.458 \,(0.0) & 0.303 \,(0.0) & -0.059 \,($1.99\times 10^{-11}$) & -0.037 \,($2.00\times 10^{-10}$) \\

\midrule

\multirow[c]{3}{*}{5}
& Statement vs Branch & 0.690 \,(0.0) & 0.496 \,(0.0) & 0.699 \,(0.0) & 0.506 \,(0.0) \\
& Statement vs MCC  & 0.839 \,(0.0) & 0.639 \,(0.0) & 0.209 \,($1.15\times 10^{-128}$) & 0.141 \,($3.99\times 10^{-129}$) \\
& Branch vs MCC     & 0.605 \,(0.0) & 0.413 \,(0.0) & -0.021 \,(0.018) & -0.008 \,(0.190) \\
\midrule

\multirow[c]{3}{*}{10}
& Statement vs Branch & 0.715 \,(0.0) & 0.517 \,(0.0) & 0.772 \,(0.0) & 0.566 \,(0.0) \\
& Statement vs MCC  & 0.878 \,(0.0) & 0.683 \,(0.0) & 0.189 \,($2.11\times 10^{-104}$) & 0.130 \,($2.87\times 10^{-110}$) \\
& Branch vs MCC     & 0.665 \,(0.0) & 0.467 \,(0.0) & 0.034 \,($8.71\times 10^{-5}$) & 0.027 \,($3.69\times 10^{-6}$) \\
\bottomrule
\end{tabular}
\end{table*}

\subsection{RQ1 Summary}
Overall, we observe several departures from the original findings:
\begin{enumerate}[leftmargin=*]
    \item \textbf{Test suite size vs.\ mutation score.} For LLM-generated tests, suite size is only weakly correlated with both raw and normalized mutation score, suggesting that generating more tests does not necessarily improve mutant-killing capability.
    \item \textbf{Coverage vs.\ mutation score without size control.} Without controlling suite size, coverage--mutation correlations are weak in both the combined and intra-model views. In the inter-model view, however, coverage becomes moderately to strongly correlated with mutation score under average aggregation, indicating that coverage is informative for comparing models.
    \item \textbf{Coverage vs.\ mutation score with size control.} After fixing suite size, correlations remain weak in the combined and intra-model views; in the inter-model view they generally decrease but often remain moderate to strong, reinforcing that coverage is more informative for model comparison than for distinguishing suites within a model.
    \item \textbf{Correlation among coverage metrics.} Correlations among coverage criteria are weaker than those reported by Inozemtseva et al.; under accumulated aggregation, modified condition coverage can be only weakly correlated with statement and branch coverage, suggesting that different coverage types are not interchangeable.
\end{enumerate}

\section{RQ2: Correlation Between Coverage and Real-Bug Detection}
In RQ2, we move beyond using mutation score as a proxy, as done by Inozemtseva et al., and analyze the correlation between coverage and real-bug detection ratio as the effectiveness measure.


\subsection{Is Test Suite Size Correlated With Real-Bug Detection?}
Unless otherwise noted, the analyses in this and the following two subsections use test suites generated from the bug-free (fixed) versions of the focal methods; the buggy-input setting is examined in Section~\ref{subsec:buggy_input_setting}. As in RQ1, we begin by examining test suite size. The overall trend is consistent in both the combined and intra-model views, where correlation coefficients remain weak. In the inter-model view, the correlation increases slightly to around $0.4$, but stays just below the moderate threshold and is not statistically significant, likely due to the small sample size. Overall, these results suggest that test suite size is not strongly correlated with real-bug detection ratio for LLM-generated tests, mirroring the weak relationship we observed for mutation score.

\begin{table}[htbp!]
\centering
\caption{Correlation between coverage and bug detection ratio across combinations of randomly selected test suites from all models; all entries are significant ($p<0.05$).}
\label{tab:cov_vs_bug_detection_ratio_combined}
\scriptsize
\setlength{\tabcolsep}{6pt}
\begin{tabular}{lcccc}
\toprule
& \multicolumn{2}{c}{Average aggregation} & \multicolumn{2}{c}{Accumulated aggregation} \\
\cmidrule(lr){2-3}\cmidrule(lr){4-5}
Coverage metric & Pearson $r$ & Kendall $\tau$ & Pearson $r$ & Kendall $\tau$ \\
\midrule
Statement & 0.368 & 0.250 & 0.473 & 0.336 \\
Branch    & 0.385 & 0.266 & 0.481 & 0.337 \\
MCC       & 0.379 & 0.261 & 0.190 & 0.127 \\
\bottomrule
\end{tabular}
\end{table}

\subsection{Is Coverage Correlated With Real-Bug Detection When Test Size Is Ignored?}
\label{subsec:cov_vs_real_bug_detect_when_size_ignored}

The overall pattern for coverage versus real-bug detection mirrors what we observed for coverage versus mutation score. In the combined view (Table~\ref{tab:cov_vs_bug_detection_ratio_combined}), coverage exhibits a weak-to-moderate correlation with bug detection effectiveness: all coefficients stay below 0.5, with accumulated modified condition coverage as the clearest outlier ($r{=}0.190$). As with the coverage--mutation analysis, switching to the intra-model view yields much weaker correlations (typically below 0.2). This drop suggests that the stronger combined-view correlation might be driven primarily by differences between models, motivating a closer look at the inter-model view.

\begin{table}[htbp!]
\centering
\caption{Inter-model correlation between coverage and bug detection ratio, with statistics aggregated into one data point per model; $p$-values are reported in parentheses.}
\label{tab:cov_vs_bug_detection_ratio_aggregated_per_model}
\scriptsize
\setlength{\tabcolsep}{6pt}
\begin{tabular}{lcccc}
\toprule
& \multicolumn{2}{c}{Average aggregation} & \multicolumn{2}{c}{Accumulated aggregation} \\
\cmidrule(lr){2-3}\cmidrule(lr){4-5}
Coverage metric & Pearson $r$ & Kendall $\tau$ & Pearson $r$ & Kendall $\tau$ \\
\midrule
Statement & 0.617 (0.025) & 0.477 (0.024) & 0.619 (0.024) & 0.555 (0.0086) \\
Branch    & 0.861 ($1.6\times 10^{-4}$) & 0.761 ($3.1\times 10^{-4}$) & 0.542 (0.056) & 0.400 (0.058) \\
MCC       & 0.573 (0.041) & 0.503 (0.017) & 0.504 (0.079) & 0.452 (0.032) \\
\bottomrule
\end{tabular}
\end{table}

We report the inter-model results in Table~\ref{tab:cov_vs_bug_detection_ratio_aggregated_per_model}. Overall, the correlations between coverage and bug detection ratio are moderate to strong, and the strongest correlations are also the most statistically significant. In particular, average branch coverage shows a strong relationship with bug detection ratio and yields the smallest $p$-values. Unlike the mutation-score results, accumulated coverage metrics also maintain moderate correlations with bug detection ratio; among them, statement coverage stands out as both statistically significant and relatively strong. Overall, these results suggest that coverage can be a useful indicator of bug detection effectiveness when comparing across different models.

\begin{table*}[htbp!]
\centering
\caption{Inter-model correlation between coverage and bug detection ratio, with statistics aggregated into one data point per model, reported for different fixed test suite sizes $k$. $p$-values are reported in parentheses.}
\label{tab:cov_vs_bug_detection_ratio_aggregated_per_model_by_k}
\scriptsize
\setlength{\tabcolsep}{6pt}
\begin{tabular}{c l cccc}
\toprule
\multirow[c]{2}{*}{$k$} & \multirow[c]{2}{*}{Coverage metric} & \multicolumn{2}{c}{Average aggregation} & \multicolumn{2}{c}{Accumulated aggregation} \\
\cmidrule(lr){3-4}\cmidrule(lr){5-6}
& & Pearson $r$ & Kendall $\tau$ & Pearson $r$ & Kendall $\tau$ \\
\midrule
\multirow[c]{3}{*}{3}
& Statement & 0.205 (0.501) & 0.222 (0.297) & 0.384 (0.195) & 0.118 (0.581) \\
& Branch    & 0.679 (0.011) & 0.405 (0.057) & 0.351 (0.240) & 0.248 (0.244) \\
& MCC       & 0.098 (0.751) & 0.118 (0.581) & -0.175 (0.568) & -0.118 (0.581) \\
\midrule
\multirow[c]{3}{*}{5}
& Statement & 0.528 (0.064) & 0.536 (0.012) & 0.690 (0.0091) & 0.484 (0.023) \\
& Branch    & 0.855 ($2.0\times 10^{-4}$) & 0.667 (0.0017) & 0.183 (0.549) & -0.013 (0.951) \\
& MCC       & 0.554 (0.050) & 0.510 (0.017) & 0.104 (0.735) & 0.144 (0.499) \\
\midrule
\multirow[c]{3}{*}{10}
& Statement & 0.589 (0.034) & 0.529 (0.012) & 0.496 (0.085) & 0.297 (0.160) \\
& Branch    & 0.858 ($1.7\times 10^{-4}$) & 0.735 ($4.9\times 10^{-4}$) & 0.353 (0.237) & 0.168 (0.427) \\
& MCC       & 0.587 (0.035) & 0.555 (0.0086) & 0.121 (0.694) & 0.142 (0.501) \\
\bottomrule
\end{tabular}
\end{table*}

\subsection{Is Coverage Correlated With Real-Bug Detection When Test Size Is Fixed?}
\label{subsec:cov_vs_real_bug_detect_when_size_fixed}
As in the size-unconstrained setting, the intra-model correlations remain weak; the combined-view correlations are consistently stronger but fall below the moderate threshold once suite size is controlled. Since these trends mirror what we observed for mutation score, we omit their detailed discussion here and focus on the inter-model analysis. Table~\ref{tab:cov_vs_bug_detection_ratio_aggregated_per_model_by_k} reports the model-wise correlations for each fixed suite size. Among the statistically significant results, correlations are consistently moderate to strong, with branch coverage standing out in particular: it is strong for $k{=}5$ and $k{=}10$, and remains just below the ``strong'' threshold for $k{=}3$ ($r{=}0.679$). Under accumulated aggregation, correlations generally drop relative to the unconstrained setting, with the only statistically significant correlation appearing between statement coverage and bug detection for $k{=}5$.

Overall, the fixed-size inter-model results reinforce our earlier finding that coverage can be a meaningful indicator of bug detection effectiveness when comparing models. The signal is strongest for average branch coverage, suggesting that different coverage types can indeed convey different information about the practical effectiveness of LLM-generated tests.

\begin{figure*}[htbp!]

\centering
\begin{minipage}{1.00\textwidth}

\begin{minipage}[t]{0.495\textwidth}
\begin{tcolorbox}[fixedbox, title=Bug-free focal method,
                  equal height group=row1]
\begin{lstlisting}[style=codeTiny,
    linebackgroundcolor={\hlRange{5}{8}}]
public static boolean equals(CharSequence cs1,
    CharSequence cs2) {
  if (cs1 == cs2)                 return true;
  if (cs1 == null || cs2 == null) return false;
  if (cs1 instanceof String && cs2 instanceof String)
    return cs1.equals(cs2);
  return CharSequenceUtils.regionMatches(cs1, false, 0,
    cs2, 0, Math.max(cs1.length(), cs2.length()));
}
\end{lstlisting}
\end{tcolorbox}
\end{minipage}%
\hfill
\begin{minipage}[t]{0.495\textwidth}
\begin{tcolorbox}[buggybox, title=Buggy focal method,
                  equal height group=row1]
\begin{lstlisting}[style=codeTiny,
    linebackgroundcolor={\hlRange{5}{5}}]
public static boolean equals(CharSequence cs1,
    CharSequence cs2) {
  if (cs1 == cs2)                 return true;
  if (cs1 == null || cs2 == null) return false;
  return cs1.equals(cs2);
}
\end{lstlisting}
\end{tcolorbox}
\end{minipage}

\vspace{1.0pt}
{\footnotesize\centering $\Downarrow$\; \emph{test generated by the LLM}\; $\Downarrow$\par}
\vspace{1.0pt}

\begin{minipage}[t]{0.495\textwidth}
\begin{tcolorbox}[fixedbox, title=Asserts the intended behavior\strut,
                  equal height group=row2]
\begin{lstlisting}[style=codeTiny, escapeinside={(*@}{@*)}]
@Test
public void testEquals_stringBuilder() {
  (*@\hltok{assertTrue}@*)(StringUtils.equals("abc",
    new StringBuilder("abc")));
}
\end{lstlisting}
\end{tcolorbox}
\end{minipage}%
\hfill
\begin{minipage}[t]{0.495\textwidth}
\begin{tcolorbox}[buggybox, title=Asserts the buggy behavior\strut,
                  equal height group=row2]
\begin{lstlisting}[style=codeTiny, escapeinside={(*@}{@*)}]
@Test
public void testEquals_stringBuilder() {
  (*@\hltok{assertFalse}@*)(StringUtils.equals("abc",
    new StringBuilder("abc")));
}
\end{lstlisting}
\end{tcolorbox}
\end{minipage}

\end{minipage}
\vspace{-0.6em}
\caption{Representative example comparing tests generated from the bug-free (left) and buggy (right) implementations of \texttt{StringUtils.equals} (Defects4J Lang-14) using the same LLM and prompt template. The top row shows the focal-method bodies provided to the model, and the bottom row shows the corresponding generated test methods. Highlighted text marks the key differences between the two implementations and between the two generated tests.}
\Description{Side-by-side comparison of the bug-free and buggy implementations of StringUtils.equals and the test an LLM generated from each; the test generated from the buggy implementation asserts the buggy behavior.}
\vspace{-0.8em}
\label{fig:buggy-vs-fixed-example}
\end{figure*}

\subsection{Does the Strong Correlation We Observe Still Hold When the Input Code Is Buggy?}
\label{subsec:buggy_input_setting}

Our inter-model analyses so far show that coverage can correlate moderately, and for some metrics even strongly, with bug detection effectiveness---but this evidence is obtained when tests are generated from bug-free code. In practice, however, the correctness of the code-under-test is not guaranteed, and the goal of testing is often to expose bugs in the buggy code itself. Prior work by Huang et al.~\cite{huang2025measuringinfluenceincorrectcode} shows that prompting LLMs with buggy code can still produce tests that expose the bug, but the number of such bug-detecting tests is substantially lower than when tests are generated from the corresponding bug-free code. A major reason is that LLMs can be misled by the buggy implementation and generate tests that assert the erroneous behavior of the buggy code rather than the intended behavior.

Figure~\ref{fig:buggy-vs-fixed-example} illustrates one such case using \texttt{StringUtils.equals} from Lang-14 in Defects4J. The method is intended to compare the contents of two \texttt{CharSequence} objects, where \texttt{CharSequence} is a Java interface implemented by common text types such as \texttt{String}, \texttt{StringBuilder}, and \texttt{StringBuffer}. Thus, \texttt{"abc"} and \texttt{new StringBuilder("abc")} should be considered equal despite having different concrete types. The bug-free implementation handles this case through \texttt{regionMatches}, whereas the buggy implementation directly calls \texttt{cs1.equals(cs2)}. Since \texttt{String.\allowbreak equals} returns \texttt{true} only when its argument is also a \texttt{String} with the same contents, it returns \texttt{false} when comparing a \texttt{String} against a \texttt{StringBuilder}. When prompted with this buggy implementation, the LLM treats this faulty behavior as intended and generates a test that asserts the incorrect outcome via \texttt{assertFalse}; in contrast, the test generated from the bug-free implementation asserts the intended behavior with \texttt{assertTrue}. This example illustrates how tests generated from buggy code can encode faulty behavior that is absent from the bug-free implementation.

Still, it remains unclear whether higher coverage on buggy code is helpful: even if an LLM treats faulty behavior as intended and thus produces weaker or incorrect assertions, exercising a broader portion of the implementation could, in principle, still increase the chance of triggering and exposing buggy behavior.

To examine this, we repeat the same correlation-analysis procedure used for bug-free code, but instead generate tests from buggy inputs and correlate coverage measured on the buggy version with bug detection effectiveness. Due to space constraints, and because the trend is consistent for all settings, we do not report the full set of coefficients here; complete statistics are available in our replication package. In summary, the correlations are weak under the combined, intra-model, and inter-model views. Although this outcome is not entirely surprising---coverage on buggy code is not necessarily informative about whether tests will detect that bug---the contrast with the bug-free setting is practically important, as it highlights a key limitation of recent LLM-based testing evaluation and test-generation practices that predominantly assume bug-free code as input.

\subsection{RQ2 Summary}
Overall, for test suites generated from bug-free code, coverage correlates more strongly with real-bug detection than with mutation score in RQ1: when comparing \emph{across} models, most coverage metrics are moderately correlated with bug detection effectiveness, and some criteria exhibit strong correlations. However, when the code-under-test is buggy, correlations between coverage and bug detection become uniformly weak across all three views, indicating that coverage measured on the buggy code is not informative of whether the generated tests detect that bug.

\section{RQ3: Correlation Between Mutation and Real-Bug Detection}
For RQ3, we leverage the data from RQ1 and RQ2 to replicate Papadakis et al.'s analysis of the relationship between mutation score and real-bug detection effectiveness for LLM-generated tests.

\subsection{Is Mutation Score Correlated With Real-Bug Detection When Test Size Is Ignored?}
We begin with the combined view (Table~\ref{tab:mut_vs_bug_detection_ratio_across_all_model}). The Pearson correlation between mutation score and bug detection ratio is moderate for average aggregation ($0.475$--$0.511$), while the corresponding Kendall $\tau$ stays below the moderate threshold; correlations are weak for accumulated aggregation. In the intra-model view, correlations drop to weak across all models and metric types (detailed results omitted for brevity). This pattern mirrors RQ1 and RQ2, suggesting that any stronger relationship, if present, is more likely to appear at the inter-model level.

\begin{table}[htbp!]
\centering
\caption{Correlation between mutation and bug detection ratio across combinations of randomly selected test suites from all models, reported for both average aggregation and accumulated aggregation; all entries are significant ($p<0.05$).}
\label{tab:mut_vs_bug_detection_ratio_across_all_model}
\scriptsize
\setlength{\tabcolsep}{6pt}
\begin{tabular}{lcccc}
\toprule
& \multicolumn{2}{c}{Average aggregation} & \multicolumn{2}{c}{Accumulated aggregation} \\
\cmidrule(lr){2-3}\cmidrule(lr){4-5}
Mutation score & Pearson $r$ & Kendall $\tau$ & Pearson $r$ & Kendall $\tau$ \\
\midrule
Raw        & 0.475 & 0.320 & 0.146 & 0.110 \\
Normalized & 0.511 & 0.373 & 0.196 & 0.123 \\
\bottomrule
\end{tabular}
\end{table}

Table~\ref{tab:mut_vs_bug_detection_ratio_aggregated_per_model} reports the inter-model results. Under both aggregation schemes, raw mutation score shows the strongest association with bug detection effectiveness: the correlation is strong under average aggregation and moderate under accumulated aggregation. This suggests that raw mutation score can serve as a useful indicator of bug detection effectiveness when comparing models. Notably, the metric variant with the strongest correlation differs from that in the combined and intra-model views. This pattern contrasts with earlier RQs, where the relative ordering of metric variants was more consistent across views, reinforcing the importance of reporting multiple metric variants.


\begin{table}[htbp!]
\centering
\caption{Inter-model correlation between mutation and bug detection ratio, with statistics aggregated into one data point per model; $p$-values are reported in parentheses.}
\label{tab:mut_vs_bug_detection_ratio_aggregated_per_model}
\scriptsize
\setlength{\tabcolsep}{6pt}
\begin{tabular}{lcccc}
\toprule
& \multicolumn{2}{c}{Average aggregation} & \multicolumn{2}{c}{Accumulated aggregation} \\
\cmidrule(lr){2-3}\cmidrule(lr){4-5}
Mutation score & Pearson $r$ & Kendall $\tau$ & Pearson $r$ & Kendall $\tau$ \\
\midrule
Raw
& 0.863 ($1.44\times 10^{-4}$) & 0.761 ($3.11\times 10^{-4}$)
& 0.622 (0.023) & 0.529 (0.012) \\
Normalized
& 0.493 (0.087) & 0.374 (0.076)
& -0.077 (0.803) & -0.039 (0.855) \\
\bottomrule
\end{tabular}
\end{table}
\vspace{-0.3cm}
\subsection{Is Mutation Score Correlated With Real-Bug Detection When Test Size Is Fixed?}
In the combined and intra-model views, the size-controlled results largely follow the same overall trend as in the size-unconstrained setting; therefore, we omit detailed coefficients for brevity and summarize the results here. In both views, correlations remain weak, with intra-model correlations consistently weaker than the combined results and average aggregation consistently stronger than accumulated aggregation. In the inter-model view (Table~\ref{tab:mut_vs_bug_detection_ratio_aggregated_per_model_fixed_k}), correlations decrease slightly after controlling for suite size but mostly remain moderate to strong and statistically significant for raw mutation score under both average and accumulated aggregation, reinforcing our observations from the previous section.


\begin{table}[htbp!]
\centering
\caption{Inter-model correlation between mutation and bug detection ratio at fixed test suite size $k$, with statistics aggregated into one data point per model; $p$-values are reported in parentheses.}
\label{tab:mut_vs_bug_detection_ratio_aggregated_per_model_fixed_k}
\scriptsize
\setlength{\tabcolsep}{6pt}
\begin{tabular}{l l cccc}
\toprule
Suite size $k$ & Mutation score & \multicolumn{2}{c}{Average aggregation} & \multicolumn{2}{c}{Accumulated aggregation} \\
\cmidrule(lr){3-4}\cmidrule(lr){5-6}
& & Pearson $r$ & Kendall $\tau$ & Pearson $r$ & Kendall $\tau$ \\
\midrule


\multirow[c]{2}{*}{3}
& Raw
& 0.702 (0.007) & 0.510 (0.017)
& 0.585 (0.036) & 0.405 (0.057) \\
& Normalized
& 0.354 (0.236) & 0.222 (0.297)
& 0.186 (0.542) & 0.013 (0.951) \\
\midrule

\multirow[c]{2}{*}{5}
& Raw
& 0.695 (0.008) & 0.510 (0.017)
& 0.533 (0.061) & 0.301 (0.158) \\
& Normalized
& 0.386 (0.192) & 0.275 (0.197)
& -0.186 (0.542) & -0.118 (0.580) \\
\midrule

\multirow[c]{2}{*}{10}
& Raw
& 0.833 ($4.05\times 10^{-4}$) & 0.658 ($1.83\times 10^{-3}$)
& 0.600 (0.030) & 0.477 (0.024) \\
& Normalized
& 0.532 (0.061) & 0.426 (0.044)
& -0.085 (0.783) & 0.013 (0.951) \\

\bottomrule
\end{tabular}
\vspace{-1.0em}
\end{table}

\subsection{RQ3 Summary}
For RQ3, we again observe a departure from the original findings of Papadakis et al. For LLM-generated tests, the correlation between mutation score and real-bug detection ratio remains at most moderate in the combined view and weak in the intra-model view, regardless of whether test suite size is controlled. In contrast, in the inter-model view, bug detection is moderately to strongly correlated with \emph{raw} mutation score under both average and accumulated aggregation, suggesting that raw mutation score can be informative for comparing bug detection effectiveness across models.

We also observe a different pattern from our earlier results: whereas differences were previously more often attributable to the choice of aggregation (average vs.\ accumulated) than to the mutation-score variant itself, in RQ3 the mutation-score variant plays a larger role. Overall, these findings motivate reporting multiple metrics under both aggregation schemes.

\vspace{-0.6em}
\section{Discussion}
Based on our findings, we summarize a set of implications, insights, and actionable recommendations that we believe can inform better practice for evaluating LLMs for test generation and for interpreting evaluation results.

\textbf{\textcircled{1}~Coverage and mutation can be informative, but only under certain conditions.}
In contrast to the conclusions of Inozemtseva et al.\ and Papadakis et al., we find that coverage and mutation can still be indicative of bug detection effectiveness, even when test suite size is controlled. However, this holds only in certain settings. When the code provided to the LLM can reasonably be assumed bug-free (e.g., regression-style scenarios where the goal is to detect bugs introduced by future changes), higher coverage and mutation scores often coincide with higher real-bug detection effectiveness when comparing \emph{across} models. In the more practical setting where the code-under-test may already be buggy and the goal is to expose that existing bug, coverage loses predictive power for bug detection effectiveness, and mutation testing is not applicable.

\textbf{\textcircled{2}~Evaluating LLMs under the practical scenario of buggy code-under-test.}
Given the weak correlations we observe between coverage and bug detection effectiveness in the buggy-code setting---together with the limited applicability of mutation analysis---we emphasize the need to directly evaluate an LLM's ability to generate tests that detect bugs in the code-under-test. This conclusion aligns with Huang et al.~\cite{huang2025measuringinfluenceincorrectcode}, who show that models' bug detection effectiveness drops substantially when the input code is buggy. Taken together, these findings motivate further research on methods that improve LLMs' ability to generate tests that expose existing bugs.

\textbf{\textcircled{3}~Different coverage criteria provide complementary signals.}
Inozemtseva et al.\ reported very strong correlations among different coverage metrics and concluded that ``Additionally, it may be in the developer's best interest to use simpler coverage measures. These measures provide a similar amount of information about the suite's effectiveness but introduce less measurement overhead.'' However, we find that, for LLM-generated tests, this conclusion does not hold. In our results, correlations between coverage criteria are only moderate to strong depending on the particular pair, and this pattern is primarily observed under the \emph{average} aggregation. Under the \emph{accumulated} aggregation, modified condition coverage correlates only weakly with statement and branch coverage. These observations suggest that reporting multiple coverage criteria, including stricter ones such as modified condition coverage, can provide additional information about LLM-generated test suites. This is also consistent with industrial practice in safety-critical domains, where standards such as DO-178B require adequate modified condition/decision coverage \cite{faa_standard}.

\textbf{\textcircled{4}~Is test suite size really a confounding factor?}
While both Inozemtseva et al.\ and Papadakis et al.\ argue that test suite size is a strong confounder in the relationships among coverage, mutation, and real-bug detection, our results for LLM-generated tests provide little support for this conclusion. We find that test suite size correlates only weakly with mutation score and with real-bug detection ratio, with at most a modest increase toward the moderate threshold in the inter-model view for bug detection. We further examine correlations between test suite size and coverage and again observe no strong relationship. Moreover, moving from the size-unconstrained to the fixed-size setting yields only a small drop in correlations: they largely retain their qualitative strength, with only occasional drops from strong to moderate. Taken together, these results suggest that test suite size is unlikely to be the primary driver of the strong correlations we observe.

\textbf{\textcircled{5}~Why do our conclusions about test suite size differ substantially from prior work?}
The weaker confounding effect of test suite size in our study likely stems from how varying-sized test suites are constructed. In our setting, each test suite is generated by the LLM for a single focal method, so differences in suite size likely reflect how many tests are needed to exercise the set of behaviors required to validate that method's expected functionality. In this sense, variation in suite size arises naturally from the testing needs of the focal method, rather than from an externally imposed target size, and therefore is less likely to exert a strong confounding effect on the other metrics. In contrast, prior studies construct suites by sampling a specified number of tests from a project-wide test pool. Under that setup, each sampled test contributes only part of the overall behavior space, so larger sampled suites mechanically aggregate more of these partial contributions and tend to achieve higher coverage, mutant killing, and bug detection simply because they include more tests, creating a stronger size-driven confounding effect.

\textbf{\textcircled{6}~How should we interpret test suite size-related metrics for LLM-generated test suites?}
Our discussion in \textcircled{4} and \textcircled{5} also informs how to interpret test-correctness measures (e.g., the number or fraction of generated tests that compile or pass), which are commonly reported in LLM test-generation evaluations~\cite{OnLLMUTGenEvalASE2024, EmpEvalforLLMUTAutoGenTSE, EvalImprChatgptUTGenFSE2024}. While producing more ``correct'' tests is desirable, correctness measures should not be conflated with bug detection effectiveness; instead, they are better viewed as measures of \emph{cost-effectiveness}. In particular, a higher fraction of compilable/passing tests indicates that less generation budget is wasted on unusable outputs.

\textbf{\textcircled{7}~What other factors may contribute to the discrepancies between our findings and prior studies?}
Beyond the difference in test suite construction discussed in \textcircled{5}, we identify two further differences that may help account for why our findings diverge from prior studies. First, LLM-based test generation differs from both human test writing and traditional automated test generation because both the behaviors to be exercised and the assertion oracles are conditioned on the prompt, focal method, and surrounding code context. By contrast, human-written tests are often guided by developers' understanding of the intended behavior, while traditional automated tools typically take the implementation as input and generate regression tests by optimizing structural testing objectives, such as coverage or mutation, over that implementation. Second, we examine correlations at multiple granularities, including combined, intra-model, and inter-model analyses, revealing that correlations can vary substantially across different perspectives, an analysis not conducted in the same way by prior studies.

\vspace{-0.6em}
\section{Threats to Validity}
\paragraph{External Validity.}
External validity concerns the generalizability of our findings beyond the studied setting. Following the scope of Inozemtseva et al., we focus on large open-source Java projects and use Defects4J, a widely adopted benchmark in both prior automated testing research and recent evaluations of LLM-based unit test generation. To reduce the risk that our results are driven by a small sample or a particular model, we conduct experiments across 11 state-of-the-art LLMs and analyze more than 100{,}000 distinct generated test cases. Nevertheless, our conclusions are limited to the Java ecosystem and the characteristics of Defects4J projects, and may not fully generalize to other programming languages or project styles. That said, the scale of our experiments should be sufficient to support the key differences we observe relative to prior studies. A second threat is that our conclusions are based on a single prompting workflow. Since identifying a universally ``best'' prompting strategy is unrealistic, we adopt a prompt template that is well justified by prior work and can be fully automated using only the code-under-test, without additional human effort, aligning with practical test-generation scenarios. We acknowledge, however, that alternative prompt designs or different contexts may yield different relationships among the studied metrics.

\paragraph{Internal Validity.}
A threat to internal validity is that some of our conclusions rely on the inter-model view, which involves a relatively small number of data points (one per model) and can reduce statistical power while increasing sensitivity to outliers. To mitigate this threat, we interpret inter-model results primarily through statistically significant coefficients and check whether the observed trends are consistent across multiple complementary settings, including randomly sampled test suites with both uncontrolled and controlled size. The fact that the same patterns persist across these settings suggests that our main observations are unlikely to be driven purely by chance.

\paragraph{Construct Validity.}
A threat to construct validity is the inherent stochasticity of LLM outputs and the dependence of results on prompting and metric choices. To reduce output variability, we set the generation temperature to $0$ to maximize determinism, while acknowledging that some nondeterminism may still arise from implementation details. Another potential threat is data contamination: developer-written tests in Defects4J may have appeared in model training data, which could inflate measured performance. While we cannot rule this out, we observe substantial differences between tests generated from buggy versus fixed versions of the same focal methods (which share the same Defects4J oracle tests), and these differences are reflected in our measured relationships among metrics; this suggests that models are responding to the provided prompt context rather than simply reproducing memorized tests.

\vspace{-0.6em}
\section{Conclusion and Future Work}
In this paper, we conduct a large-scale replication study of Inozemtseva et al.\ and Papadakis et al.\ to examine how widely used proxy metrics---test coverage and mutation score---relate to the real-bug detection effectiveness of LLM-generated test suites. Our conclusions differ substantially from those of the two prior studies. In particular, we find that the usefulness of coverage and mutation is highly context-dependent. When the code provided to the LLM can be reasonably assumed bug-free, certain forms of coverage and mutation provide meaningful signals when comparing \emph{across} models. In contrast, in the practically common scenario where the code-under-test may already be buggy and the goal is to expose that bug, coverage becomes unreliable and mutation analysis is not applicable. We also find little evidence that test suite size is a dominant confounder for correlations among coverage, mutation, and real-bug detection for LLM-generated tests. Taken together, these results help interpret prior findings and motivate future evaluations to measure what matters most: whether generated tests can reveal bugs in the code-under-test.

Our study also suggests several directions for future work. First, prior LLM-based test generation studies adopt a wide range of prompt templates and prompting strategies; a natural next step is to systematically vary prompt construction (e.g., the amount and type of code context, formatting, and auxiliary instructions) and examine how these choices affect the observed relationships among coverage, mutation, and real-bug detection effectiveness. Second, many recent techniques improve proxy outcomes such as compilation success, pass rate, coverage, or mutation score through specialized prompting or post-processing; it remains unclear when such improvements translate into higher \emph{real-bug detection effectiveness}, and clarifying this linkage would help determine when optimizing these proxies is justified. Third, our experiments adopt a predominantly white-box prompting setting in which the code-under-test is provided; future work should assess whether our observations hold in black-box or specification-based settings, where models generate tests from natural-language specifications, docstrings, or other behavioral descriptions without access to the implementation. Fourth, our study focuses on the overall correlations among coverage, mutation, and real-bug detection effectiveness for LLM-generated Java tests, but these relationships may vary across projects; investigating such project-specific correlations is therefore an important direction for future work. Finally, future work should compare tests from different sources---LLM-generated, human-written, and non-LLM test automation tools---to understand how these relationships differ across settings and what drives those differences.

\begin{acks}
This work was supported by the Natural Sciences and Engineering Research Council of Canada (NSERC) under Grant RGPIN-2022-04154.
\end{acks}

\section*{Data Availability}
The data used in this study is drawn from the publicly available Defects4J dataset~\citep{Defects4j}. To support reproducibility, we provide a replication package~\citep{zhao2026covmutreplication} containing our test-generation code, data-preprocessing scripts, and the raw data used in our analysis. All results reported in the paper can be reproduced using this package. The package is publicly available on GitHub at \url{https://github.com/drixs2050/Cov_mut_bug_detect_correlation} and permanently archived on Zenodo (DOI: \href{https://doi.org/10.5281/zenodo.21429528}{10.5281/zenodo.21429528}).


\bibliographystyle{ACM-Reference-Format}
\bibliography{ISSTA-2026-replicability}


\begin{thebibliography}{61}


\ifx \showCODEN    \undefined \def \showCODEN     #1{\unskip}     \fi
\ifx \showISBNx    \undefined \def \showISBNx     #1{\unskip}     \fi
\ifx \showISBNxiii \undefined \def \showISBNxiii  #1{\unskip}     \fi
\ifx \showISSN     \undefined \def \showISSN      #1{\unskip}     \fi
\ifx \showLCCN     \undefined \def \showLCCN      #1{\unskip}     \fi
\ifx \shownote     \undefined \def \shownote      #1{#1}          \fi
\ifx \showarticletitle \undefined \def \showarticletitle #1{#1}   \fi
\ifx \showURL      \undefined \def \showURL       {\relax}        \fi
\providecommand\bibfield[2]{#2}
\providecommand\bibinfo[2]{#2}
\providecommand\natexlab[1]{#1}
\providecommand\showeprint[2][]{arXiv:#2}

\bibitem[cod(2025)]%
        {codecover}
 \bibinfo{year}{2025}\natexlab{}.
\newblock \bibinfo{title}{CodeCover}.
\newblock \bibinfo{howpublished}{\url{http://codecover.org/}}.
\newblock
\shownote{Accessed: 2025-09-10}.
\newblock


\bibitem[Abdullin et~al\mbox{.}(2025)]%
        {abdullin2025testwarscomparativestudy}
\bibfield{author}{\bibinfo{person}{Azat Abdullin}, \bibinfo{person}{Pouria Derakhshanfar}, {and} \bibinfo{person}{Annibale Panichella}.} \bibinfo{year}{2025}\natexlab{}.
\newblock \bibinfo{title}{Test Wars: A Comparative Study of SBST, Symbolic Execution, and LLM-Based Approaches to Unit Test Generation}.
\newblock
\showeprint[arxiv]{2501.10200}~[cs.SE]
\href{https://doi.org/10.48550/arXiv.2501.10200}{doi:\nolinkurl{10.48550/arXiv.2501.10200}}


\bibitem[Alagarsamy et~al\mbox{.}(2024)]%
        {A3TestIST}
\bibfield{author}{\bibinfo{person}{Saranya Alagarsamy}, \bibinfo{person}{Chakkrit Tantithamthavorn}, {and} \bibinfo{person}{Aldeida Aleti}.} \bibinfo{year}{2024}\natexlab{}.
\newblock \showarticletitle{A3Test: Assertion-Augmented Automated Test Case Generation}.
\newblock \bibinfo{journal}{\emph{Information and Software Technology}}  \bibinfo{volume}{176} (\bibinfo{year}{2024}), \bibinfo{pages}{107565}.
\newblock
\showISSN{0950-5849}
\href{https://doi.org/10.1016/j.infsof.2024.107565}{doi:\nolinkurl{10.1016/j.infsof.2024.107565}}


\bibitem[Andrews et~al\mbox{.}(2005)]%
        {corr_mut_bug_2}
\bibfield{author}{\bibinfo{person}{J.~H. Andrews}, \bibinfo{person}{L.~C. Briand}, {and} \bibinfo{person}{Y. Labiche}.} \bibinfo{year}{2005}\natexlab{}.
\newblock \showarticletitle{Is mutation an appropriate tool for testing experiments?}. In \bibinfo{booktitle}{\emph{Proceedings of the 27th International Conference on Software Engineering}} (St. Louis, MO, USA) \emph{(\bibinfo{series}{ICSE '05})}. \bibinfo{publisher}{Association for Computing Machinery}, \bibinfo{address}{New York, NY, USA}, \bibinfo{pages}{402–411}.
\newblock
\showISBNx{1581139632}
\href{https://doi.org/10.1145/1062455.1062530}{doi:\nolinkurl{10.1145/1062455.1062530}}


\bibitem[Anthropic(2025)]%
        {anthropic2025claude4sonnet}
\bibfield{author}{\bibinfo{person}{Anthropic}.} \bibinfo{year}{2025}\natexlab{}.
\newblock \bibinfo{title}{Claude 4 Sonnet}.
\newblock \bibinfo{howpublished}{\url{https://www.anthropic.com/claude/sonnet}}.
\newblock
\shownote{Accessed: 2025-05-23}.
\newblock


\bibitem[Austin et~al\mbox{.}(2021)]%
        {mbpp}
\bibfield{author}{\bibinfo{person}{Jacob Austin}, \bibinfo{person}{Augustus Odena}, \bibinfo{person}{Maxwell Nye}, \bibinfo{person}{Maarten Bosma}, \bibinfo{person}{Henryk Michalewski}, \bibinfo{person}{David Dohan}, \bibinfo{person}{Ellen Jiang}, \bibinfo{person}{Carrie Cai}, \bibinfo{person}{Michael Terry}, \bibinfo{person}{Quoc Le}, {and} \bibinfo{person}{Charles Sutton}.} \bibinfo{year}{2021}\natexlab{}.
\newblock \bibinfo{title}{Program Synthesis with Large Language Models}.
\newblock
\showeprint[arxiv]{2108.07732}~[cs.PL]
\href{https://doi.org/10.48550/arXiv.2108.07732}{doi:\nolinkurl{10.48550/arXiv.2108.07732}}


\bibitem[Bhatia et~al\mbox{.}(2024)]%
        {UTGenUsingGenAIICSEWorkShop2024}
\bibfield{author}{\bibinfo{person}{Shreya Bhatia}, \bibinfo{person}{Tarushi Gandhi}, \bibinfo{person}{Dhruv Kumar}, {and} \bibinfo{person}{Pankaj Jalote}.} \bibinfo{year}{2024}\natexlab{}.
\newblock \showarticletitle{Unit Test Generation using Generative AI: A Comparative Performance Analysis of Autogeneration Tools}. In \bibinfo{booktitle}{\emph{Proceedings of the 1st International Workshop on Large Language Models for Code}} (Lisbon, Portugal) \emph{(\bibinfo{series}{LLM4Code '24})}. \bibinfo{publisher}{Association for Computing Machinery}, \bibinfo{address}{New York, NY, USA}, \bibinfo{pages}{54–61}.
\newblock
\showISBNx{9798400705793}
\href{https://doi.org/10.1145/3643795.3648396}{doi:\nolinkurl{10.1145/3643795.3648396}}


\bibitem[Brunsfeld(2018)]%
        {tree-sitter}
\bibfield{author}{\bibinfo{person}{Max Brunsfeld}.} \bibinfo{year}{2018}\natexlab{}.
\newblock \bibinfo{title}{Tree-sitter: An incremental parsing system for programming tools}.
\newblock
\shownote{Accessed: 2025-02-21}.
\newblock
\href{https://doi.org/10.5281/zenodo.4619183}{doi:\nolinkurl{10.5281/zenodo.4619183}}


\bibitem[Cai and Lyu(2005)]%
        {corr_cov_bug_human_3}
\bibfield{author}{\bibinfo{person}{Xia Cai} {and} \bibinfo{person}{Michael~R. Lyu}.} \bibinfo{year}{2005}\natexlab{}.
\newblock \showarticletitle{The effect of code coverage on fault detection under different testing profiles}.
\newblock \bibinfo{journal}{\emph{SIGSOFT Softw. Eng. Notes}} \bibinfo{volume}{30}, \bibinfo{number}{4} (\bibinfo{date}{July} \bibinfo{year}{2005}), \bibinfo{pages}{1–7}.
\newblock
\showISSN{0163-5948}
\href{https://doi.org/10.1145/1082983.1083288}{doi:\nolinkurl{10.1145/1082983.1083288}}


\bibitem[Carver et~al\mbox{.}(2014)]%
        {Concept_rep_def_2}
\bibfield{author}{\bibinfo{person}{Jeffrey~C. Carver}, \bibinfo{person}{Natalia Juristo}, \bibinfo{person}{Maria~Teresa Baldassarre}, {and} \bibinfo{person}{Sira Vegas}.} \bibinfo{year}{2014}\natexlab{}.
\newblock \showarticletitle{Replications of software engineering experiments}.
\newblock \bibinfo{journal}{\emph{Empirical Software Engineering}} \bibinfo{volume}{19}, \bibinfo{number}{2} (\bibinfo{year}{2014}), \bibinfo{pages}{267--276}.
\newblock
\href{https://doi.org/10.1007/s10664-013-9290-8}{doi:\nolinkurl{10.1007/s10664-013-9290-8}}


\bibitem[Chen et~al\mbox{.}(2021)]%
        {HumanEval}
\bibfield{author}{\bibinfo{person}{Mark Chen} {et~al\mbox{.}}} \bibinfo{year}{2021}\natexlab{}.
\newblock \bibinfo{title}{Evaluating Large Language Models Trained on Code}.
\newblock
\showeprint[arxiv]{2107.03374}~[cs.LG]
\href{https://doi.org/10.48550/arXiv.2107.03374}{doi:\nolinkurl{10.48550/arXiv.2107.03374}}


\bibitem[Chen et~al\mbox{.}(2024)]%
        {ChatUniTestFSE2024Demo}
\bibfield{author}{\bibinfo{person}{Yinghao Chen}, \bibinfo{person}{Zehao Hu}, \bibinfo{person}{Chen Zhi}, \bibinfo{person}{Junxiao Han}, \bibinfo{person}{Shuiguang Deng}, {and} \bibinfo{person}{Jianwei Yin}.} \bibinfo{year}{2024}\natexlab{}.
\newblock \showarticletitle{ChatUniTest: A Framework for LLM-Based Test Generation}. In \bibinfo{booktitle}{\emph{Companion Proceedings of the 32nd ACM International Conference on the Foundations of Software Engineering}} (Porto de Galinhas, Brazil) \emph{(\bibinfo{series}{FSE 2024})}. \bibinfo{publisher}{Association for Computing Machinery}, \bibinfo{address}{New York, NY, USA}, \bibinfo{pages}{572–576}.
\newblock
\showISBNx{9798400706585}
\href{https://doi.org/10.1145/3663529.3663801}{doi:\nolinkurl{10.1145/3663529.3663801}}


\bibitem[Coles et~al\mbox{.}(2016)]%
        {pit}
\bibfield{author}{\bibinfo{person}{Henry Coles}, \bibinfo{person}{Thomas Laurent}, \bibinfo{person}{Christopher Henard}, \bibinfo{person}{Mike Papadakis}, {and} \bibinfo{person}{Anthony Ventresque}.} \bibinfo{year}{2016}\natexlab{}.
\newblock \showarticletitle{PIT: a practical mutation testing tool for Java (demo)}. In \bibinfo{booktitle}{\emph{Proceedings of the 25th International Symposium on Software Testing and Analysis}} (Saarbr\"{u}cken, Germany) \emph{(\bibinfo{series}{ISSTA 2016})}. \bibinfo{publisher}{Association for Computing Machinery}, \bibinfo{address}{New York, NY, USA}, \bibinfo{pages}{449–452}.
\newblock
\showISBNx{9781450343909}
\href{https://doi.org/10.1145/2931037.2948707}{doi:\nolinkurl{10.1145/2931037.2948707}}


\bibitem[Dakhel et~al\mbox{.}(2024)]%
        {TestGenLLMandMutIST}
\bibfield{author}{\bibinfo{person}{Arghavan~Moradi Dakhel}, \bibinfo{person}{Amin Nikanjam}, \bibinfo{person}{Vahid Majdinasab}, \bibinfo{person}{Foutse Khomh}, {and} \bibinfo{person}{Michel~C. Desmarais}.} \bibinfo{year}{2024}\natexlab{}.
\newblock \showarticletitle{Effective test generation using pre-trained Large Language Models and mutation testing}.
\newblock \bibinfo{journal}{\emph{Information and Software Technology}}  \bibinfo{volume}{171} (\bibinfo{year}{2024}), \bibinfo{pages}{107468}.
\newblock
\showISSN{0950-5849}
\href{https://doi.org/10.1016/j.infsof.2024.107468}{doi:\nolinkurl{10.1016/j.infsof.2024.107468}}


\bibitem[{DeepSeek-AI}(2025a)]%
        {deepseekai2025deepseekr1incentivizingreasoningcapability}
\bibfield{author}{\bibinfo{person}{{DeepSeek-AI}}.} \bibinfo{year}{2025}\natexlab{a}.
\newblock \bibinfo{title}{DeepSeek-R1: Incentivizing Reasoning Capability in LLMs via Reinforcement Learning}.
\newblock
\showeprint[arxiv]{2501.12948}~[cs.CL]
\href{https://doi.org/10.48550/arXiv.2501.12948}{doi:\nolinkurl{10.48550/arXiv.2501.12948}}


\bibitem[{DeepSeek-AI}(2025b)]%
        {deepseekai2025deepseekv3technicalreport}
\bibfield{author}{\bibinfo{person}{{DeepSeek-AI}}.} \bibinfo{year}{2025}\natexlab{b}.
\newblock \bibinfo{title}{DeepSeek-V3 Technical Report}.
\newblock
\showeprint[arxiv]{2412.19437}~[cs.CL]
\href{https://doi.org/10.48550/arXiv.2412.19437}{doi:\nolinkurl{10.48550/arXiv.2412.19437}}


\bibitem[Frankl and Iakounenko(1998)]%
        {corr_cov_bug_human_2}
\bibfield{author}{\bibinfo{person}{Phyllis~G. Frankl} {and} \bibinfo{person}{Oleg Iakounenko}.} \bibinfo{year}{1998}\natexlab{}.
\newblock \showarticletitle{Further empirical studies of test effectiveness}. In \bibinfo{booktitle}{\emph{Proceedings of the 6th ACM SIGSOFT International Symposium on Foundations of Software Engineering}} (Lake Buena Vista, Florida, USA) \emph{(\bibinfo{series}{SIGSOFT '98/FSE-6})}. \bibinfo{publisher}{Association for Computing Machinery}, \bibinfo{address}{New York, NY, USA}, \bibinfo{pages}{153–162}.
\newblock
\showISBNx{1581131089}
\href{https://doi.org/10.1145/288195.288298}{doi:\nolinkurl{10.1145/288195.288298}}


\bibitem[Fraser and Arcuri(2014)]%
        {Evosuite}
\bibfield{author}{\bibinfo{person}{Gordon Fraser} {and} \bibinfo{person}{Andrea Arcuri}.} \bibinfo{year}{2014}\natexlab{}.
\newblock \showarticletitle{A Large-Scale Evaluation of Automated Unit Test Generation Using EvoSuite}.
\newblock \bibinfo{journal}{\emph{ACM Trans. Softw. Eng. Methodol.}} \bibinfo{volume}{24}, \bibinfo{number}{2}, Article \bibinfo{articleno}{8} (\bibinfo{date}{Dec.} \bibinfo{year}{2014}), \bibinfo{numpages}{42}~pages.
\newblock
\showISSN{1049-331X}
\href{https://doi.org/10.1145/2685612}{doi:\nolinkurl{10.1145/2685612}}


\bibitem[Fraser et~al\mbox{.}(2013)]%
        {whitebox_test}
\bibfield{author}{\bibinfo{person}{Gordon Fraser}, \bibinfo{person}{Matt Staats}, \bibinfo{person}{Phil McMinn}, \bibinfo{person}{Andrea Arcuri}, {and} \bibinfo{person}{Frank Padberg}.} \bibinfo{year}{2013}\natexlab{}.
\newblock \showarticletitle{Does automated white-box test generation really help software testers?}. In \bibinfo{booktitle}{\emph{Proceedings of the 2013 International Symposium on Software Testing and Analysis}} (Lugano, Switzerland) \emph{(\bibinfo{series}{ISSTA 2013})}. \bibinfo{publisher}{Association for Computing Machinery}, \bibinfo{address}{New York, NY, USA}, \bibinfo{pages}{291–301}.
\newblock
\showISBNx{9781450321594}
\href{https://doi.org/10.1145/2483760.2483774}{doi:\nolinkurl{10.1145/2483760.2483774}}


\bibitem[Gligoric et~al\mbox{.}(2013)]%
        {corr_cov_bug_human_4}
\bibfield{author}{\bibinfo{person}{Milos Gligoric}, \bibinfo{person}{Alex Groce}, \bibinfo{person}{Chaoqiang Zhang}, \bibinfo{person}{Rohan Sharma}, \bibinfo{person}{Mohammad~Amin Alipour}, {and} \bibinfo{person}{Darko Marinov}.} \bibinfo{year}{2013}\natexlab{}.
\newblock \showarticletitle{Comparing non-adequate test suites using coverage criteria}. In \bibinfo{booktitle}{\emph{Proceedings of the 2013 International Symposium on Software Testing and Analysis}} (Lugano, Switzerland) \emph{(\bibinfo{series}{ISSTA 2013})}. \bibinfo{publisher}{Association for Computing Machinery}, \bibinfo{address}{New York, NY, USA}, \bibinfo{pages}{302–313}.
\newblock
\showISBNx{9781450321594}
\href{https://doi.org/10.1145/2483760.2483769}{doi:\nolinkurl{10.1145/2483760.2483769}}


\bibitem[Google(2025a)]%
        {Google2025Gemini2.5Flash}
\bibfield{author}{\bibinfo{person}{Google}.} \bibinfo{year}{2025}\natexlab{a}.
\newblock \bibinfo{title}{Gemini 2.5 Flash Model}.
\newblock \bibinfo{howpublished}{\url{https://ai.google.dev/gemini-api/docs/models\#gemini-2.5-flash}}.
\newblock
\shownote{Accessed: 2025-08-19}.
\newblock


\bibitem[Google(2025b)]%
        {Google2025Gemini2.5Pro}
\bibfield{author}{\bibinfo{person}{Google}.} \bibinfo{year}{2025}\natexlab{b}.
\newblock \bibinfo{title}{Gemini 2.5 Pro Model}.
\newblock \bibinfo{howpublished}{\url{https://ai.google.dev/gemini-api/docs/models\#gemini-2.5-pro}}.
\newblock
\shownote{Accessed: 2025-08-19}.
\newblock


\bibitem[Guilford(1956)]%
        {guilfordscale}
\bibfield{author}{\bibinfo{person}{J.P. Guilford}.} \bibinfo{year}{1956}\natexlab{}.
\newblock \bibinfo{booktitle}{\emph{Fundamental Statistics in Psychology and Education}}.
\newblock \bibinfo{publisher}{McGraw-Hill}.
\newblock
\showLCCN{gb56008458}
\urldef\tempurl%
\url{https://books.google.ca/books?id=u-G10ZqLhtsC}
\showURL{%
\tempurl}


\bibitem[Guilherme and Vincenzi(2023)]%
        {InitialInvestigationChatGPT}
\bibfield{author}{\bibinfo{person}{Vitor Guilherme} {and} \bibinfo{person}{Auri Vincenzi}.} \bibinfo{year}{2023}\natexlab{}.
\newblock \showarticletitle{An initial investigation of ChatGPT unit test generation capability}. In \bibinfo{booktitle}{\emph{Proceedings of the 8th Brazilian Symposium on Systematic and Automated Software Testing}} (Campo Grande, MS, Brazil) \emph{(\bibinfo{series}{SAST '23})}. \bibinfo{publisher}{Association for Computing Machinery}, \bibinfo{address}{New York, NY, USA}, \bibinfo{pages}{15–24}.
\newblock
\showISBNx{9798400716294}
\href{https://doi.org/10.1145/3624032.3624035}{doi:\nolinkurl{10.1145/3624032.3624035}}


\bibitem[Hayhurst and Veerhusen(2001)]%
        {faa_standard}
\bibfield{author}{\bibinfo{person}{Kelly~J. Hayhurst} {and} \bibinfo{person}{Dan~S. Veerhusen}.} \bibinfo{year}{2001}\natexlab{}.
\newblock \showarticletitle{A practical approach to modified condition/decision coverage}. In \bibinfo{booktitle}{\emph{Proceedings of the 20th Digital Avionics Systems Conference (DASC)}}, Vol.~\bibinfo{volume}{1}. IEEE, \bibinfo{pages}{1B2/1--1B2/10}.
\newblock
\href{https://doi.org/10.1109/DASC.2001.963305}{doi:\nolinkurl{10.1109/DASC.2001.963305}}


\bibitem[Hendrycks et~al\mbox{.}(2021)]%
        {apps}
\bibfield{author}{\bibinfo{person}{Dan Hendrycks}, \bibinfo{person}{Steven Basart}, \bibinfo{person}{Saurav Kadavath}, \bibinfo{person}{Mantas Mazeika}, \bibinfo{person}{Akul Arora}, \bibinfo{person}{Ethan Guo}, \bibinfo{person}{Collin Burns}, \bibinfo{person}{Samir Puranik}, \bibinfo{person}{Horace He}, \bibinfo{person}{Dawn Song}, {and} \bibinfo{person}{Jacob Steinhardt}.} \bibinfo{year}{2021}\natexlab{}.
\newblock \showarticletitle{Measuring Coding Challenge Competence With APPS}. In \bibinfo{booktitle}{\emph{Advances in Neural Information Processing Systems (Datasets and Benchmarks Track)}}.
\newblock
\href{https://doi.org/10.48550/arXiv.2105.09938}{doi:\nolinkurl{10.48550/arXiv.2105.09938}}


\bibitem[Hossain et~al\mbox{.}(2025)]%
        {Doc2OracLL}
\bibfield{author}{\bibinfo{person}{Soneya~Binta Hossain}, \bibinfo{person}{Raygan Taylor}, {and} \bibinfo{person}{Matthew Dwyer}.} \bibinfo{year}{2025}\natexlab{}.
\newblock \showarticletitle{Doc2OracLL: Investigating the Impact of Documentation on LLM-Based Test Oracle Generation}.
\newblock \bibinfo{journal}{\emph{Proc. ACM Softw. Eng.}} \bibinfo{volume}{2}, \bibinfo{number}{FSE}, Article \bibinfo{articleno}{FSE084} (\bibinfo{date}{June} \bibinfo{year}{2025}), \bibinfo{numpages}{22}~pages.
\newblock
\href{https://doi.org/10.1145/3729354}{doi:\nolinkurl{10.1145/3729354}}


\bibitem[Hu et~al\mbox{.}(2023)]%
        {Ceprot}
\bibfield{author}{\bibinfo{person}{Xing Hu}, \bibinfo{person}{Zhuang Liu}, \bibinfo{person}{Xin Xia}, \bibinfo{person}{Zhongxin Liu}, \bibinfo{person}{Tongtong Xu}, {and} \bibinfo{person}{Xiaohu Yang}.} \bibinfo{year}{2023}\natexlab{}.
\newblock \showarticletitle{Identify and Update Test Cases When Production Code Changes: A Transformer-Based Approach}. In \bibinfo{booktitle}{\emph{Proceedings of the 38th IEEE/ACM International Conference on Automated Software Engineering}} (Echternach, Luxembourg) \emph{(\bibinfo{series}{ASE '23})}. \bibinfo{publisher}{IEEE Press}, \bibinfo{pages}{1111–1122}.
\newblock
\showISBNx{9798350329964}
\href{https://doi.org/10.1109/ASE56229.2023.00165}{doi:\nolinkurl{10.1109/ASE56229.2023.00165}}


\bibitem[Huang et~al\mbox{.}(2025)]%
        {huang2025measuringinfluenceincorrectcode}
\bibfield{author}{\bibinfo{person}{Dong Huang}, \bibinfo{person}{Jie~M. Zhang}, \bibinfo{person}{Mark Harman}, \bibinfo{person}{Mingzhe Du}, {and} \bibinfo{person}{Heming Cui}.} \bibinfo{year}{2025}\natexlab{}.
\newblock \bibinfo{title}{Measuring the Influence of Incorrect Code on Test Generation}.
\newblock
\showeprint[arxiv]{2409.09464}~[cs.SE]
\href{https://doi.org/10.48550/arXiv.2409.09464}{doi:\nolinkurl{10.48550/arXiv.2409.09464}}


\bibitem[Inozemtseva and Holmes(2014)]%
        {CovNotStronglyCorrWithEffICSE2014}
\bibfield{author}{\bibinfo{person}{Laura Inozemtseva} {and} \bibinfo{person}{Reid Holmes}.} \bibinfo{year}{2014}\natexlab{}.
\newblock \showarticletitle{Coverage is not strongly correlated with test suite effectiveness}. In \bibinfo{booktitle}{\emph{Proceedings of the 36th International Conference on Software Engineering}} (Hyderabad, India) \emph{(\bibinfo{series}{ICSE 2014})}. \bibinfo{publisher}{Association for Computing Machinery}, \bibinfo{address}{New York, NY, USA}, \bibinfo{pages}{435–445}.
\newblock
\showISBNx{9781450327565}
\href{https://doi.org/10.1145/2568225.2568271}{doi:\nolinkurl{10.1145/2568225.2568271}}


\bibitem[Jain et~al\mbox{.}(2025)]%
        {jain2025testgeneval}
\bibfield{author}{\bibinfo{person}{Kush Jain}, \bibinfo{person}{Gabriel Synnaeve}, {and} \bibinfo{person}{Baptiste Roziere}.} \bibinfo{year}{2025}\natexlab{}.
\newblock \showarticletitle{TestGenEval: A Real World Unit Test Generation and Test Completion Benchmark}. In \bibinfo{booktitle}{\emph{The Thirteenth International Conference on Learning Representations}}.
\newblock
\href{https://doi.org/10.48550/arXiv.2410.00752}{doi:\nolinkurl{10.48550/arXiv.2410.00752}}


\bibitem[Jimenez et~al\mbox{.}(2024)]%
        {jimenez2024swebench}
\bibfield{author}{\bibinfo{person}{Carlos~E. Jimenez}, \bibinfo{person}{John Yang}, \bibinfo{person}{Alexander Wettig}, \bibinfo{person}{Shunyu Yao}, \bibinfo{person}{Kexin Pei}, \bibinfo{person}{Ofir Press}, {and} \bibinfo{person}{Karthik~R. Narasimhan}.} \bibinfo{year}{2024}\natexlab{}.
\newblock \showarticletitle{{SWE}-bench: Can Language Models Resolve Real-World GitHub Issues?}. In \bibinfo{booktitle}{\emph{The Twelfth International Conference on Learning Representations}}.
\newblock
\href{https://doi.org/10.48550/arXiv.2310.06770}{doi:\nolinkurl{10.48550/arXiv.2310.06770}}


\bibitem[Jorgensen(2013)]%
        {testing_textbook}
\bibfield{author}{\bibinfo{person}{Paul~C. Jorgensen}.} \bibinfo{year}{2013}\natexlab{}.
\newblock \bibinfo{booktitle}{\emph{Software Testing: A Craftsman's Approach} (\bibinfo{edition}{fourth} ed.)}.
\newblock \bibinfo{publisher}{Auerbach Publications}.
\newblock
\showISBNx{9781466560703}
\href{https://doi.org/10.1201/b15980}{doi:\nolinkurl{10.1201/b15980}}


\bibitem[Just et~al\mbox{.}(2014a)]%
        {Defects4j}
\bibfield{author}{\bibinfo{person}{Ren\'{e} Just}, \bibinfo{person}{Darioush Jalali}, {and} \bibinfo{person}{Michael~D. Ernst}.} \bibinfo{year}{2014}\natexlab{a}.
\newblock \showarticletitle{Defects4J: a database of existing faults to enable controlled testing studies for Java programs}. In \bibinfo{booktitle}{\emph{Proceedings of the 2014 International Symposium on Software Testing and Analysis}} (San Jose, CA, USA) \emph{(\bibinfo{series}{ISSTA 2014})}. \bibinfo{publisher}{Association for Computing Machinery}, \bibinfo{address}{New York, NY, USA}, \bibinfo{pages}{437–440}.
\newblock
\showISBNx{9781450326452}
\href{https://doi.org/10.1145/2610384.2628055}{doi:\nolinkurl{10.1145/2610384.2628055}}


\bibitem[Just et~al\mbox{.}(2014b)]%
        {corr_mut_bug_1}
\bibfield{author}{\bibinfo{person}{Ren\'{e} Just}, \bibinfo{person}{Darioush Jalali}, \bibinfo{person}{Laura Inozemtseva}, \bibinfo{person}{Michael~D. Ernst}, \bibinfo{person}{Reid Holmes}, {and} \bibinfo{person}{Gordon Fraser}.} \bibinfo{year}{2014}\natexlab{b}.
\newblock \showarticletitle{Are mutants a valid substitute for real faults in software testing?}. In \bibinfo{booktitle}{\emph{Proceedings of the 22nd ACM SIGSOFT International Symposium on Foundations of Software Engineering}} (Hong Kong, China) \emph{(\bibinfo{series}{FSE 2014})}. \bibinfo{publisher}{Association for Computing Machinery}, \bibinfo{address}{New York, NY, USA}, \bibinfo{pages}{654–665}.
\newblock
\showISBNx{9781450330565}
\href{https://doi.org/10.1145/2635868.2635929}{doi:\nolinkurl{10.1145/2635868.2635929}}


\bibitem[Lops et~al\mbox{.}(2025)]%
        {sysforautocreateandassess}
\bibfield{author}{\bibinfo{person}{Andrea Lops}, \bibinfo{person}{Fedelucio Narducci}, \bibinfo{person}{Azzurra Ragone}, \bibinfo{person}{Michelantonio Trizio}, {and} \bibinfo{person}{Claudio Bartolini}.} \bibinfo{year}{2025}\natexlab{}.
\newblock \showarticletitle{A System for Automated Unit Test Generation using Large Language Models and Assessment of Generated Test Suites}. In \bibinfo{booktitle}{\emph{2025 IEEE International Conference on Software Testing, Verification and Validation Workshops (ICSTW)}}. \bibinfo{publisher}{IEEE}, \bibinfo{pages}{29–36}.
\newblock
\href{https://doi.org/10.1109/icstw64639.2025.10962454}{doi:\nolinkurl{10.1109/icstw64639.2025.10962454}}


\bibitem[OpenAI(2025a)]%
        {openai2025gpt41}
\bibfield{author}{\bibinfo{person}{OpenAI}.} \bibinfo{year}{2025}\natexlab{a}.
\newblock \bibinfo{title}{GPT-4.1}.
\newblock \bibinfo{howpublished}{\url{https://openai.com/index/gpt-4-1/}}.
\newblock
\shownote{Accessed: 2025-08-19}.
\newblock


\bibitem[OpenAI(2025b)]%
        {openai2025o4mini}
\bibfield{author}{\bibinfo{person}{OpenAI}.} \bibinfo{year}{2025}\natexlab{b}.
\newblock \bibinfo{title}{OpenAI Models - O4 Mini}.
\newblock \bibinfo{howpublished}{\url{https://platform.openai.com/docs/models/o4-mini}}.
\newblock
\shownote{Accessed: 2025-08-19}.
\newblock


\bibitem[Panichella et~al\mbox{.}(2018)]%
        {dynamosa}
\bibfield{author}{\bibinfo{person}{Annibale Panichella}, \bibinfo{person}{Fitsum~Meshesha Kifetew}, {and} \bibinfo{person}{Paolo Tonella}.} \bibinfo{year}{2018}\natexlab{}.
\newblock \showarticletitle{Automated Test Case Generation as a Many-Objective Optimisation Problem with Dynamic Selection of the Targets}.
\newblock \bibinfo{journal}{\emph{IEEE Transactions on Software Engineering}} \bibinfo{volume}{44}, \bibinfo{number}{2} (\bibinfo{year}{2018}), \bibinfo{pages}{122--158}.
\newblock
\href{https://doi.org/10.1109/TSE.2017.2663435}{doi:\nolinkurl{10.1109/TSE.2017.2663435}}


\bibitem[Papadakis et~al\mbox{.}(2018)]%
        {papadakis_mutation_vs_real_bug_detect}
\bibfield{author}{\bibinfo{person}{Mike Papadakis}, \bibinfo{person}{Donghwan Shin}, \bibinfo{person}{Shin Yoo}, {and} \bibinfo{person}{Doo-Hwan Bae}.} \bibinfo{year}{2018}\natexlab{}.
\newblock \showarticletitle{Are mutation scores correlated with real fault detection? A large scale empirical study on the relationship between mutants and real faults}. In \bibinfo{booktitle}{\emph{Proceedings of the 40th International Conference on Software Engineering}} (Gothenburg, Sweden) \emph{(\bibinfo{series}{ICSE '18})}. \bibinfo{publisher}{Association for Computing Machinery}, \bibinfo{address}{New York, NY, USA}, \bibinfo{pages}{537–548}.
\newblock
\showISBNx{9781450356381}
\href{https://doi.org/10.1145/3180155.3180183}{doi:\nolinkurl{10.1145/3180155.3180183}}


\bibitem[{Qwen Team}(2025a)]%
        {alibaba2025qwen3coder}
\bibfield{author}{\bibinfo{person}{{Qwen Team}}.} \bibinfo{year}{2025}\natexlab{a}.
\newblock \bibinfo{title}{Qwen3-Coder: Agentic Coding in the World}.
\newblock \bibinfo{howpublished}{\url{https://qwenlm.github.io/blog/qwen3-coder/}}.
\newblock
\shownote{Accessed: 2025-08-19}.
\newblock


\bibitem[{Qwen Team}(2025b)]%
        {alibaba2025qwen3}
\bibfield{author}{\bibinfo{person}{{Qwen Team}}.} \bibinfo{year}{2025}\natexlab{b}.
\newblock \bibinfo{title}{Qwen3: Think Deeper, Act Faster}.
\newblock \bibinfo{howpublished}{\url{https://qwenlm.github.io/blog/qwen3/}}.
\newblock
\shownote{Accessed: 2025-08-19}.
\newblock


\bibitem[Rao et~al\mbox{.}(2023)]%
        {CAT-LMASE2023}
\bibfield{author}{\bibinfo{person}{Nikitha Rao}, \bibinfo{person}{Kush Jain}, \bibinfo{person}{Uri Alon}, \bibinfo{person}{Claire~Le Goues}, {and} \bibinfo{person}{Vincent~J. Hellendoorn}.} \bibinfo{year}{2023}\natexlab{}.
\newblock \showarticletitle{CAT-LM Training Language Models on Aligned Code and Tests}. In \bibinfo{booktitle}{\emph{Proceedings of the 38th IEEE/ACM International Conference on Automated Software Engineering}} (Echternach, Luxembourg) \emph{(\bibinfo{series}{ASE '23})}. \bibinfo{publisher}{IEEE Press}, \bibinfo{pages}{409–420}.
\newblock
\showISBNx{9798350329964}
\href{https://doi.org/10.1109/ASE56229.2023.00193}{doi:\nolinkurl{10.1109/ASE56229.2023.00193}}


\bibitem[Rojas et~al\mbox{.}(2015)]%
        {automatedunittestevosuiteregress}
\bibfield{author}{\bibinfo{person}{Jos\'{e}~Miguel Rojas}, \bibinfo{person}{Gordon Fraser}, {and} \bibinfo{person}{Andrea Arcuri}.} \bibinfo{year}{2015}\natexlab{}.
\newblock \showarticletitle{Automated unit test generation during software development: a controlled experiment and think-aloud observations}. In \bibinfo{booktitle}{\emph{Proceedings of the 2015 International Symposium on Software Testing and Analysis}} (Baltimore, MD, USA) \emph{(\bibinfo{series}{ISSTA 2015})}. \bibinfo{publisher}{Association for Computing Machinery}, \bibinfo{address}{New York, NY, USA}, \bibinfo{pages}{338–349}.
\newblock
\showISBNx{9781450336208}
\href{https://doi.org/10.1145/2771783.2771801}{doi:\nolinkurl{10.1145/2771783.2771801}}


\bibitem[Schäfer et~al\mbox{.}(2024)]%
        {EmpEvalforLLMUTAutoGenTSE}
\bibfield{author}{\bibinfo{person}{Max Schäfer}, \bibinfo{person}{Sarah Nadi}, \bibinfo{person}{Aryaz Eghbali}, {and} \bibinfo{person}{Frank Tip}.} \bibinfo{year}{2024}\natexlab{}.
\newblock \showarticletitle{An Empirical Evaluation of Using Large Language Models for Automated Unit Test Generation}.
\newblock \bibinfo{journal}{\emph{IEEE Transactions on Software Engineering}} \bibinfo{volume}{50}, \bibinfo{number}{1} (\bibinfo{year}{2024}), \bibinfo{pages}{85--105}.
\newblock
\href{https://doi.org/10.1109/TSE.2023.3334955}{doi:\nolinkurl{10.1109/TSE.2023.3334955}}


\bibitem[Shang et~al\mbox{.}(2025)]%
        {LargeScaleEmpStudyFTUTGenISSTA2025}
\bibfield{author}{\bibinfo{person}{Ye Shang}, \bibinfo{person}{Quanjun Zhang}, \bibinfo{person}{Chunrong Fang}, \bibinfo{person}{Siqi Gu}, \bibinfo{person}{Jianyi Zhou}, {and} \bibinfo{person}{Zhenyu Chen}.} \bibinfo{year}{2025}\natexlab{}.
\newblock \showarticletitle{A Large-Scale Empirical Study on Fine-Tuning Large Language Models for Unit Testing}.
\newblock \bibinfo{journal}{\emph{Proc. ACM Softw. Eng.}} \bibinfo{volume}{2}, \bibinfo{number}{ISSTA}, Article \bibinfo{articleno}{ISSTA074} (\bibinfo{date}{June} \bibinfo{year}{2025}), \bibinfo{numpages}{23}~pages.
\newblock
\href{https://doi.org/10.1145/3728951}{doi:\nolinkurl{10.1145/3728951}}


\bibitem[Shull et~al\mbox{.}(2008)]%
        {Concept_rep_def_1}
\bibfield{author}{\bibinfo{person}{Forrest~J. Shull}, \bibinfo{person}{Jeffrey~C. Carver}, \bibinfo{person}{Sira Vegas}, {and} \bibinfo{person}{Natalia Juristo}.} \bibinfo{year}{2008}\natexlab{}.
\newblock \showarticletitle{The role of replications in Empirical Software Engineering}.
\newblock \bibinfo{journal}{\emph{Empirical Softw. Engg.}} \bibinfo{volume}{13}, \bibinfo{number}{2} (\bibinfo{date}{April} \bibinfo{year}{2008}), \bibinfo{pages}{211–218}.
\newblock
\showISSN{1382-3256}
\href{https://doi.org/10.1007/s10664-008-9060-1}{doi:\nolinkurl{10.1007/s10664-008-9060-1}}


\bibitem[Siddiq et~al\mbox{.}(2024)]%
        {LLMforJunitTestGenEASE2024}
\bibfield{author}{\bibinfo{person}{Mohammed~Latif Siddiq}, \bibinfo{person}{Joanna~Cecilia Da~Silva~Santos}, \bibinfo{person}{Ridwanul~Hasan Tanvir}, \bibinfo{person}{Noshin Ulfat}, \bibinfo{person}{Fahmid Al~Rifat}, {and} \bibinfo{person}{Vin\'{\i}cius Carvalho~Lopes}.} \bibinfo{year}{2024}\natexlab{}.
\newblock \showarticletitle{Using Large Language Models to Generate JUnit Tests: An Empirical Study}. In \bibinfo{booktitle}{\emph{Proceedings of the 28th International Conference on Evaluation and Assessment in Software Engineering}} (Salerno, Italy) \emph{(\bibinfo{series}{EASE '24})}. \bibinfo{publisher}{Association for Computing Machinery}, \bibinfo{address}{New York, NY, USA}, \bibinfo{pages}{313–322}.
\newblock
\showISBNx{9798400717017}
\href{https://doi.org/10.1145/3661167.3661216}{doi:\nolinkurl{10.1145/3661167.3661216}}


\bibitem[Silva et~al\mbox{.}(2024)]%
        {Gitbug}
\bibfield{author}{\bibinfo{person}{Andr\'e Silva}, \bibinfo{person}{Nuno Saavedra}, {and} \bibinfo{person}{Martin Monperrus}.} \bibinfo{year}{2024}\natexlab{}.
\newblock \showarticletitle{{GitBug-Java: A Reproducible Benchmark of Recent Java Bugs}}. In \bibinfo{booktitle}{\emph{2024 IEEE/ACM 21st International Conference on Mining Software Repositories (MSR)}}. \bibinfo{publisher}{IEEE Computer Society}, \bibinfo{address}{Los Alamitos, CA, USA}, \bibinfo{pages}{118--122}.
\newblock
\href{https://doi.org/10.1145/3643991.3644884}{doi:\nolinkurl{10.1145/3643991.3644884}}


\bibitem[Tang et~al\mbox{.}(2024)]%
        {ChatVSSBSTTSE}
\bibfield{author}{\bibinfo{person}{Yutian Tang}, \bibinfo{person}{Zhijie Liu}, \bibinfo{person}{Zhichao Zhou}, {and} \bibinfo{person}{Xiapu Luo}.} \bibinfo{year}{2024}\natexlab{}.
\newblock \showarticletitle{ChatGPT vs SBST: A Comparative Assessment of Unit Test Suite Generation}.
\newblock \bibinfo{journal}{\emph{IEEE Transactions on Software Engineering}} \bibinfo{volume}{50}, \bibinfo{number}{6} (\bibinfo{year}{2024}), \bibinfo{pages}{1340--1359}.
\newblock
\href{https://doi.org/10.1109/TSE.2024.3382365}{doi:\nolinkurl{10.1109/TSE.2024.3382365}}


\bibitem[Tufano et~al\mbox{.}(2021)]%
        {UnitTestCaseGenTrans}
\bibfield{author}{\bibinfo{person}{Michele Tufano}, \bibinfo{person}{Dawn Drain}, \bibinfo{person}{Alexey Svyatkovskiy}, \bibinfo{person}{Shao~Kun Deng}, {and} \bibinfo{person}{Neel Sundaresan}.} \bibinfo{year}{2021}\natexlab{}.
\newblock \bibinfo{title}{Unit Test Case Generation with Transformers and Focal Context}.
\newblock
\showeprint[arxiv]{2009.05617}~[cs.SE]
\href{https://doi.org/10.48550/arXiv.2009.05617}{doi:\nolinkurl{10.48550/arXiv.2009.05617}}


\bibitem[Wang et~al\mbox{.}(2025)]%
        {wang-etal-2025-testeval}
\bibfield{author}{\bibinfo{person}{Wenhan Wang}, \bibinfo{person}{Chenyuan Yang}, \bibinfo{person}{Zhijie Wang}, \bibinfo{person}{Yuheng Huang}, \bibinfo{person}{Zhaoyang Chu}, \bibinfo{person}{Da Song}, \bibinfo{person}{Lingming Zhang}, \bibinfo{person}{An~Ran Chen}, {and} \bibinfo{person}{Lei Ma}.} \bibinfo{year}{2025}\natexlab{}.
\newblock \showarticletitle{{TestEval}: Benchmarking Large Language Models for Test Case Generation}. In \bibinfo{booktitle}{\emph{Findings of the Association for Computational Linguistics: NAACL 2025}}, \bibfield{editor}{\bibinfo{person}{Luis Chiruzzo}, \bibinfo{person}{Alan Ritter}, {and} \bibinfo{person}{Lu~Wang}} (Eds.). \bibinfo{publisher}{Association for Computational Linguistics}, \bibinfo{address}{Albuquerque, New Mexico}, \bibinfo{pages}{3547--3562}.
\newblock
\showISBNx{979-8-89176-195-7}
\href{https://doi.org/10.18653/v1/2025.findings-naacl.197}{doi:\nolinkurl{10.18653/v1/2025.findings-naacl.197}}


\bibitem[Watson et~al\mbox{.}(2020)]%
        {ATLAS}
\bibfield{author}{\bibinfo{person}{Cody Watson}, \bibinfo{person}{Michele Tufano}, \bibinfo{person}{Kevin Moran}, \bibinfo{person}{Gabriele Bavota}, {and} \bibinfo{person}{Denys Poshyvanyk}.} \bibinfo{year}{2020}\natexlab{}.
\newblock \showarticletitle{On learning meaningful assert statements for unit test cases}. In \bibinfo{booktitle}{\emph{Proceedings of the ACM/IEEE 42nd International Conference on Software Engineering}} (Seoul, South Korea) \emph{(\bibinfo{series}{ICSE '20})}. \bibinfo{publisher}{Association for Computing Machinery}, \bibinfo{address}{New York, NY, USA}, \bibinfo{pages}{1398–1409}.
\newblock
\showISBNx{9781450371216}
\href{https://doi.org/10.1145/3377811.3380429}{doi:\nolinkurl{10.1145/3377811.3380429}}


\bibitem[Widyasari et~al\mbox{.}(2020)]%
        {BugsInPy}
\bibfield{author}{\bibinfo{person}{Ratnadira Widyasari}, \bibinfo{person}{Sheng~Qin Sim}, \bibinfo{person}{Camellia Lok}, \bibinfo{person}{Haodi Qi}, \bibinfo{person}{Jack Phan}, \bibinfo{person}{Qijin Tay}, \bibinfo{person}{Constance Tan}, \bibinfo{person}{Fiona Wee}, \bibinfo{person}{Jodie~Ethelda Tan}, \bibinfo{person}{Yuheng Yieh}, \bibinfo{person}{Brian Goh}, \bibinfo{person}{Ferdian Thung}, \bibinfo{person}{Hong~Jin Kang}, \bibinfo{person}{Thong Hoang}, \bibinfo{person}{David Lo}, {and} \bibinfo{person}{Eng~Lieh Ouh}.} \bibinfo{year}{2020}\natexlab{}.
\newblock \showarticletitle{BugsInPy: a database of existing bugs in Python programs to enable controlled testing and debugging studies}. In \bibinfo{booktitle}{\emph{Proceedings of the 28th ACM Joint Meeting on European Software Engineering Conference and Symposium on the Foundations of Software Engineering}} (Virtual Event, USA) \emph{(\bibinfo{series}{ESEC/FSE 2020})}. \bibinfo{publisher}{Association for Computing Machinery}, \bibinfo{address}{New York, NY, USA}, \bibinfo{pages}{1556–1560}.
\newblock
\showISBNx{9781450370431}
\href{https://doi.org/10.1145/3368089.3417943}{doi:\nolinkurl{10.1145/3368089.3417943}}


\bibitem[xAI(2025a)]%
        {xai2025grok3}
\bibfield{author}{\bibinfo{person}{xAI}.} \bibinfo{year}{2025}\natexlab{a}.
\newblock \bibinfo{title}{Grok-3 Model Documentation}.
\newblock \bibinfo{howpublished}{\url{https://docs.x.ai/docs/models/grok-3}}.
\newblock
\shownote{Accessed: 2025-08-19}.
\newblock


\bibitem[xAI(2025b)]%
        {xai2025grok4}
\bibfield{author}{\bibinfo{person}{xAI}.} \bibinfo{year}{2025}\natexlab{b}.
\newblock \bibinfo{title}{Grok-4 Model Documentation}.
\newblock \bibinfo{howpublished}{\url{https://docs.x.ai/docs/models/grok-4}}.
\newblock
\shownote{Accessed: 2025-08-19}.
\newblock


\bibitem[Yang et~al\mbox{.}(2024)]%
        {OnLLMUTGenEvalASE2024}
\bibfield{author}{\bibinfo{person}{Lin Yang}, \bibinfo{person}{Chen Yang}, \bibinfo{person}{Shutao Gao}, \bibinfo{person}{Weijing Wang}, \bibinfo{person}{Bo Wang}, \bibinfo{person}{Qihao Zhu}, \bibinfo{person}{Xiao Chu}, \bibinfo{person}{Jianyi Zhou}, \bibinfo{person}{Guangtai Liang}, \bibinfo{person}{Qianxiang Wang}, {and} \bibinfo{person}{Junjie Chen}.} \bibinfo{year}{2024}\natexlab{}.
\newblock \showarticletitle{On the Evaluation of Large Language Models in Unit Test Generation}. In \bibinfo{booktitle}{\emph{Proceedings of the 39th IEEE/ACM International Conference on Automated Software Engineering}} (Sacramento, CA, USA) \emph{(\bibinfo{series}{ASE '24})}. \bibinfo{publisher}{Association for Computing Machinery}, \bibinfo{address}{New York, NY, USA}, \bibinfo{pages}{1607–1619}.
\newblock
\showISBNx{9798400712487}
\href{https://doi.org/10.1145/3691620.3695529}{doi:\nolinkurl{10.1145/3691620.3695529}}


\bibitem[Yuan et~al\mbox{.}(2024)]%
        {EvalImprChatgptUTGenFSE2024}
\bibfield{author}{\bibinfo{person}{Zhiqiang Yuan}, \bibinfo{person}{Mingwei Liu}, \bibinfo{person}{Shiji Ding}, \bibinfo{person}{Kaixin Wang}, \bibinfo{person}{Yixuan Chen}, \bibinfo{person}{Xin Peng}, {and} \bibinfo{person}{Yiling Lou}.} \bibinfo{year}{2024}\natexlab{}.
\newblock \showarticletitle{Evaluating and Improving ChatGPT for Unit Test Generation}.
\newblock \bibinfo{journal}{\emph{Proc. ACM Softw. Eng.}} \bibinfo{volume}{1}, \bibinfo{number}{FSE}, Article \bibinfo{articleno}{76} (\bibinfo{date}{July} \bibinfo{year}{2024}), \bibinfo{numpages}{24}~pages.
\newblock
\href{https://doi.org/10.1145/3660783}{doi:\nolinkurl{10.1145/3660783}}


\bibitem[Zhang et~al\mbox{.}(2025)]%
        {unittestgenerationreview}
\bibfield{author}{\bibinfo{person}{Quanjun Zhang}, \bibinfo{person}{Chunrong Fang}, \bibinfo{person}{Siqi Gu}, \bibinfo{person}{Ye Shang}, \bibinfo{person}{Zhenyu Chen}, {and} \bibinfo{person}{Liang Xiao}.} \bibinfo{year}{2025}\natexlab{}.
\newblock \bibinfo{title}{Large Language Models for Unit Testing: A Systematic Literature Review}.
\newblock
\showeprint[arxiv]{2506.15227}~[cs.SE]
\href{https://doi.org/10.48550/arXiv.2506.15227}{doi:\nolinkurl{10.48550/arXiv.2506.15227}}


\bibitem[Zhang et~al\mbox{.}(2024)]%
        {zhang2024testbenchevaluatingclassleveltest}
\bibfield{author}{\bibinfo{person}{Quanjun Zhang}, \bibinfo{person}{Ye Shang}, \bibinfo{person}{Chunrong Fang}, \bibinfo{person}{Siqi Gu}, \bibinfo{person}{Jianyi Zhou}, {and} \bibinfo{person}{Zhenyu Chen}.} \bibinfo{year}{2024}\natexlab{}.
\newblock \bibinfo{title}{TestBench: Evaluating Class-Level Test Case Generation Capability of Large Language Models}.
\newblock
\showeprint[arxiv]{2409.17561}~[cs.SE]
\href{https://doi.org/10.48550/arXiv.2409.17561}{doi:\nolinkurl{10.48550/arXiv.2409.17561}}


\bibitem[Zhao et~al\mbox{.}(2026)]%
        {zhao2026covmutreplication}
\bibfield{author}{\bibinfo{person}{Junda Zhao}, \bibinfo{person}{Shurui Zhou}, {and} \bibinfo{person}{Eldan Cohen}.} \bibinfo{year}{2026}\natexlab{}.
\newblock \bibinfo{title}{Replication Package for ``Do Coverage and Mutation Scores of {LLM}-Generated Test Suites Correlate With Their Effectiveness? (Replicability Study)''}.
\newblock \bibinfo{howpublished}{\url{https://github.com/drixs2050/Cov_mut_bug_detect_correlation}}.
\newblock
\href{https://doi.org/10.5281/zenodo.21437945}{doi:\nolinkurl{10.5281/zenodo.21437945}}


\end{thebibliography}










\end{document}